\documentclass[twocolumn]{aastex631}
\usepackage{soul}
\usepackage{amsmath}

\definecolor{myblue}{rgb}{0.,0.4,0.7}
\definecolor{mygreen}{rgb}{0.,0.8,0.3}
\definecolor{myred}{rgb}{1.0,0.4,0}

\shorttitle{Environmental Differences in Dwarf Galaxies}
\shortauthors{Christensen et al.}

\usepackage{abrev} 
\usepackage{sidecap}
\usepackage{color}

\begin{document}

\title{Environment Matters: Predicted Differences in the Stellar Mass\-Halo Mass Relation and History of Star Formation for Dwarf Galaxies}

\author{Charlotte R. Christensen}
\affil{Physics Department, Grinnell College, 1116 Eighth Ave., Grinnell, IA 50112, United States}
\email{christenc@grinnell.edu}

\author{Alyson Brooks}
\affil{Department of Physics and Astronomy, Rutgers University, the State University of New Jersey, 136 Frelinghuysen Road, Piscataway, NJ 08854-8019, United States}
\affil{Center for Computational Astrophysics, Flatiron Institute, 162 Fifth Avenue, New York, NY 10010, USA}

\author{Ferah Munshi}
\affil{Department of Physics and Astronomy, George Mason University, 4400 University Drive, MSN: 3F3,
Fairfax, VA 22030, United States}

\author{Claire Riggs}
\affil{Department of Physics and Astronomy, Rutgers University, the State University of New Jersey, 136 Frelinghuysen Road, Piscataway, NJ 08854-8019, United States}

\author{Jordan Van Nest}
\affil{Homer L. Dodge Department of Physics and Astronomy,
The University of Oklahoma
440 W.~Brooks St.,
Norman, OK 73019, United States}

\author{Hollis Akins}
\affil{Department of Astronomy, The University of Texas at Austin
2515 Speedway, Stop C1400
Austin, Texas 78712-1205, United States}

\author{Thomas R Quinn}
\affil{Department of Astronomy, University of Washington, 3910 15$^{th}$ Ave.~NE,
Seattle, WA 98195-1700, United States}

\author{Lucas Chamberland}
\affil{Physics Department, Grinnell College, 1116 Eighth Ave., Grinnell, IA 50112, United States}

\def\path{./}

\begin{abstract}
We are entering an era in which we will be able to detect and characterize hundreds of dwarf galaxies within the Local Volume.  
It is already known that a strong dichotomy exists in the gas content and star formation properties of field dwarf galaxies versus satellite dwarfs of larger galaxies. In this work, we study the more subtle differences that may be detectable in galaxies as a function of distance from a massive galaxy, such as the Milky Way. We compare smoothed particle hydrodynamic simulations of dwarf galaxies formed in a Local Volume-like environment (several Mpc away from a massive galaxy) to those formed nearer to Milky Way-mass halos. We find that the impact of environment on dwarf galaxies extends even beyond the immediate region surrounding Milky Way-mass halos. Even before being accreted as satellites, dwarf galaxies near a Milky Way-mass halo tend to have higher stellar masses for their halo mass than more isolated galaxies. Dwarf galaxies in high-density environments also tend to grow faster and form their stars earlier. We show observational predictions that demonstrate how these trends manifest in lower quenching rates, higher \textsc{H\,i} fractions, and bluer colors for more isolated dwarf galaxies.
\end{abstract}

\section{Introduction}\label{sec:intro}

The next few years are poised to lead to the discovery of hundreds of dwarf galaxies within the Local Volume.  First, the Vera C.~Rubin Observatory's Legacy Survey of Space \& Time (LSST) is expected to discover hundreds of galaxies down into the ultra-faint dwarf range within a few Mpc \citep{Tollerud2008, Simon2019}.  Second, WALLABY, an Australian SKA Pathfinder survey that is already producing data in its pilot phase, will push the \textsc{H\,i} detection threshold down to $10^5$ M$_{\odot}$ within a few Mpc, and down to $10^6$ M$_{\odot}$ within the Local Volume.  This opens the potential to discover low-mass galaxies via their \textsc{H\,i} content, as has already been demonstrated in the ALFALFA survey by the discovery of galaxies such as Leo P \citep{Giovanelli2013} and those in the Survey of \textsc{H\,i} in Extremely Low-Mass Dwarfs \citep[SHIELD;][]{Cannon2011, McQuinn2021}.  These discoveries will allow an unprecedented characterization of the properties of low-luminosity galaxies in the Local Volume.  In this paper, we explore the expected properties of simulated dwarf galaxies as a function of distance from a massive galaxy, which can be directly tested by these future observations. 

It is already established that there is a remarkable dichotomy between the star formation rates of dwarf satellite galaxies and field dwarfs.  Nearly all of the Milky Way's and M31's known satellites are dwarf spheroidal galaxies (dSphs) with quenched star formation, minimal \textsc{H\,i} gas, and elliptical morphologies \citep{Spekkens2014, Putman2021}.  And yet galaxies in the Local Volume and beyond appear to be gas-rich and not quenched \citep{Geha2012, McConnachie2012}.

Clearly, dwarf galaxy evolution is strongly influenced in the presence of a massive halo.  To what distance the massive halo environment influences evolution remains an open question.

There are multiple physical processes which may impact dwarf galaxy evolution near a massive halo.
For example, some apparently-isolated dwarf galaxies may be ``backsplash" systems \citep{Gill2005} whose evolution has been affected through previous encounters with a massive system \citep{Teyssier2012, simpson_quenching_2018}.
This previous interaction history can have profound effects on the structure and star formation histories of the dwarf galaxies.
\citet{Buck2019} demonstrated that substantial (often times $>$ 80\%) mass loss may occur in backsplash galaxies during their pericentric passage.
Such backsplash systems have similar stellar masses to their field compatriots but lower virial masses, resulting in lower mass-to-light ratios \citep{Knebe2011} and lower stellar velocity dispersions \citep{Applebaum2021}.
Since most quenching occurs on rapid timescales ($\lesssim$ 1 Gyr), the quenched fractions of such backsplash galaxies more closely resemble satellites than field galaxies \citep{simpson_quenching_2018}. 
In fact, \citet{Wetzel2014} argued that the entirety of the environmental dependency of galaxy quenching could be explained if backsplash galaxies share the same star formation histories as satellite galaxies.

Additionally, the star formation histories of satellites may have been shaped prior to accretion by interactions with similar mass galaxies \citep{Zabludoff1998}.  The interactions could lead to increased consumption of gas \citep{Deason2014}.
In eventual satellites, this process is known as ``pre-processing" \citep[e.g.,][]{Wetzel2013, Wetzel2015, simpson_quenching_2018, Samuel2022} but the effects of repeated interactions of galaxies in denser environments are likely universal.

Finally, galaxies in denser environments may also influence each other by shaping the chemical and thermodynamics properties of the surrounding gas.
For example, \citet{Arora2022} showed that central dwarf galaxies in simulated Local Group environments had greater amounts of cold gas and metals than their isolated analogues.
They hypothesize that metal enrichment from the more massive halos results in greater rates of gas cooling and increased early star formation.
Similarly, the increased UV background expected in denser environments may play a similar role by changing the local reionization history.
Simulations that allow for inhomogeneous reionization show that halos that experience later reionization have higher baryonic fractions than those of similar masses that experienced earlier reionization \citep{Katz2020}, which could allow for dwarf galaxies in low density environments to have higher $M_*/M_{\mathrm{vir}}$.

However, there is also evidence that the environmental effects on the star formation histories of dwarf galaxies may not be limited to direct galaxy-galaxy interactions such as those described above.
For example, \citet{Gallart2015} divides the local group dwarf galaxies into ``fast dwarfs," whose star formation history is dominated by an early, short ($\lesssim$ a few Gyr) star formation event, and ``slow dwarfs" that have experienced more continuous star formation.
The type of star formation history a galaxy experiences appears to be most closely related to the location of the dwarf, with galaxies thought to have formed at larger distances from the Milky Way more likely to be ``slow dwarfs" than more nearby galaxies \citep[e.g., WLM, DDO 210, and Leo A;][]{Albers2019, Cole2007, Cole2014}.
\citet{Gallart2015} argues that the different star formation histories are a signal of the characteristic density within which the galaxy formed.
The argument is that in higher density systems, galaxies with masses just above the threshold for star formation during reionization collapse early and so have their star formation truncated by the combined effects of reionization and stellar feedback.
In contrast, a galaxy with a mass just above the threshold mass in a low-density environment may collapse more slowly and the resulting delay in star formation allows it to continue to later times.

This understanding of earlier star formation within higher density environments is consistent with the model of hierarchical galaxy formation.
In hierarchical galaxy formation, material within higher-$\sigma$ density fluctuations undergoes gravitational collapse earlier.
Material from these high-$\sigma$ peaks is then more likely to be found close to the center of massive structures \citep{Bardeen1986, Diemand2005, Moore2006}.
We would, therefore, expect a correspondence between the strength of the initial density fluctuation, the time of gravitational collapse, and the proximity to a massive galaxy. 
Both observational \citep{Poudel2017} and computational \citep{Xu2020} comparisons of central galaxies formed within different cosmic web environments have indicated that galaxies formed in richer environments have higher stellar masses and lower specific star formation rates.
When considering the effect on lower-mass galaxies in a Milky Way environment, 
\citet{Brooks12a} found that satellite dSph galaxies assembled their virial masses earlier than field galaxies of similar luminosities.
\citet{Garrison-Kimmel2019} also found that satellites with $M_* < 10^7 \Msun$ formed their stars earlier than central galaxies of equivalent stellar mass near Milky Way-mass halos, and that both populations formed stars earlier than ``highly isolated dwarf galaxies."
Intriguingly, \citet{Garrison-Kimmel2019} further found some evidence that central galaxies in a Local Group environment both formed their stars earlier and assembled their peak virial mass earlier than those near a single Milky Way-mass host; the authors remain agnostic as to whether this difference is the result of dark matter accretion history or gravitational influences from the more massive companions.

Semianalytic models have verified that the stochasticity in the mass assembly histories of galaxies with present day virial masses of $10^9 - 10^{10} \Msun$ can lead to qualitatively different star formation histories, whose differences are greatly amplified by reionization \citep{Ledinauskas2018}.
Smoothed particle hydrodynamic simulations of low-mass galaxies with the same final virial mass but different formation timescales have reinforced this view.
\citet{Fitts2017} found that the earliest forming halos with the highest concentration in the dark matter only version of the simulation obtained the highest stellar mass.
\citet{Rey2019} showed that the ability of the earlier-growing halo to form more stars before reionization resulted in a greater stellar-to-virial-mass ratio.
Similarly, \citet{Sawala2016} showed that the later collapse of halos in lower-density regions resulted in them being less likely to be luminous.
The possibility of a galaxy quenched by reionization reigniting at a later time is also highly dependent on the assembly history---\citet{Rey2020} showed earlier post-reionization mass accretion was more efficient at reigniting star formation.

Simulations that link star formation to environment have seen similar patterns when comparing dwarf galaxies formed in filaments to voids.
For example, \citet{Liao2019} find that the dwarf galaxy halos formed in filaments have higher baryon and stellar fractions at $z \sim 2.5$ than their field counterparts, indicating that these filaments helped funnel gas into the galaxy. 
This finding was reproduced in \citet{Zheng2021} for simulations with stellar feedback, which also found that at $z \sim 2.5$, dwarfs in filaments tended to have both bluer colors and more rapid star formation than their field counterparts.
In comparison, \citet{Xu2020} examined galaxies with M$_{200}$ h$^{-1} > 10^{10} \Msun$ from the {\sc Eagle} simulations, and found no difference in the specific star formation rates at $z \geq 1$ and that the differences at $z$ = 0 could be entirely attributed to higher quenched fractions in denser environments.  

In this paper, we first explore the resulting stellar mass--halo mass (SMHM) relation in simulated dwarf galaxies as a function of environment. In general, the SMHM relation for rich environments depends strongly on how the virial mass is defined. 
For example, when the SMHM relation is measured directly for the IllustrisTNG simulations  \citep{Engler2020}, the reduction of the dynamical mass through tidal stripping instead moves the relation to the left (i.e., higher $M_*/M_{\mathrm{vir}}$) for satellite galaxies \citep[see also][]{Sawala2015, Buck2019, Tremmel2020, Munshi2021}.
For this reason, most analysis of the SMHM for simulations is computed using either the peak virial mass \citep[e.g.][]{Munshi2021} or the virial mass prior to infall \citep[e.g.][]{Read2017}, in order to remove the effect of tidal disruption of the halo.
Yet even when the impact of tidal disruption is removed, any or all of the processes described above may impact the star formation, and thus could plausibly affect the resulting SMHM relation.

In this work we compare simulations of dwarf galaxies formed near Milky Way-mass halos to those formed in more isolated environments ( $>$ 1 Mpc from a Milky Way-mass galaxy).  We find that star formation varies as a function environment in the simulations.  This impacts not only the SMHM relation, but also additional properties (e.g., star formation histories, gas content, color) that can be directly tested as a function of environment.
\S\ref{sec:SMHM} compares the SMHM relation, \S\ref{sec:BMHM} compares the baryonic mass--halo mass relation, \S\ref{sec:sfh} compares the star formation histories, while \S\ref{sec:mass_assem} examines the mass assembly histories.
\S\ref{sec:evol} examines the evolution of the SMHM relation, while \S\ref{sec:theory} compares against previous theoretical work.  
\S\ref{sec:obs} illustrates the observational comparisons and predictions.

\section{Methods}\label{sec:meth}

\begin{deluxetable*}{c cccc cccc}
\tablenum{1}
\tablecaption{Description of Simulations\label{tab:sims}}
\tablewidth{0pt}
\tablehead{
\colhead{Name} & 
\colhead{Cosmology} & 
\colhead{$\epsilon$} & \colhead{m$_{DM}$} & \colhead{m$_{gas}$} &
\colhead{$c^*$} & \colhead{$T_{SF}$} &\colhead{$\rho_{SF}$} & \colhead{$\epsilon_{SN}$} \\
\colhead{} &
\colhead{}  & 
\colhead{[pc]}  & \colhead{[$10^3 \Msun$]} & \colhead{[$10^3 \Msun$]} &
\colhead{}  & \colhead{[K]} &\colhead{[m$_p$ cm$^{-3}$]} &\colhead{[ergs SN$^{-1}$]} 
}
\startdata
MARVEL$^{1, 2}$ & WMAP 3   &
60  & 6.66  &   1.41   & 
$0.1 f_{\mathrm{H}_2}$ & $10^3$ & 0.1 & $1.5 \times 10^{51}$ \\
JL Mint$^{3}$ & Planck 2015  &
87  & 17.9  & 3.31  & 
$0.1 f_{\mathrm{H}_2}$ & $10^3$ & 0.1 & $1.5 \times 10^{51}$ \\
JL Near Mint$^{1, 3, 4}$ & Planck 2015  &
170 & 42 & 27 & 
$0.1 f_{\mathrm{H}_2}$ & $10^3$ & 0.1 & $1.5 \times 10^{51}$ \\
R\sc{omulus}25$^{5}$ & Planck 2015  &
250 & 339   & 212  &
0.15& $10^4$ & 0.2 &
$0.75 \times 10^{51}$  \\ 
\enddata
\tablecomments{
Simulations are included in 1: \citet{Bellovary2018}; 2: \citet{Munshi2021}; 3: \citet{Applebaum2021}; 4: \citet{Akins2021}; 5: \citet{Tremmel2017}.
}
\end{deluxetable*}

The simulated dwarf galaxies analyzed in this paper were collected from three different suites of cosmological simulations spanning different environments, resolutions, and sub-grid physical recipes.
These suites of simulations are summarized in Table~\ref{tab:sims}, and described in greater detail in \S~\ref{sec:ICs}.
Briefly, though, the `{\sc Marvel}-ous Dwarfs' suite consists of simulations that model collections of field dwarf galaxies formed in the low-density environment of a cosmic sheet, similar to the Local Volume environment in regards to distance to a massive galaxy. 
In contrast, the `{\sc D.\,C.\,Justice League}' suite models the rich environment around Milky Way-mass halos.
The {\sc Romulus25} simulation provides a lower-resolution bridge across both environments by simulating a uniform (25 Mpc)$^3$ cosmological volume.

\subsection{Initial Conditions}\label{sec:ICs}
The {\sc Marvel}-ous Dwarfs and {\sc D.\,C.\,Justice League} simulation suites are a set of eight (four {\sc Marvel}-ous Dwarfs, four {\sc D.\,C.\,Justice League}) zoom-in simulations \citep{katz93}.
This zoom-in technique ensures that the region of interest is simulated at high resolution while also including effects from the broader cosmological environment.
In this technique, galaxies were pre-selected from a uniform volume N-body simulation, and the source region for these galaxies was resimulated at high resolution with full hydrodynamics.

The {\sc Marvel}-ous Dwarfs simulations are each focused on a cosmic sheet containing a collection of dwarf galaxies.
These simulations use cosmological parameters from WMAP 3 \citep{Spergel07}.
Together, these volumes contain 68 dwarf galaxies with at least 100 particles, our resolution limit for identifying a galaxy (table~\ref{tab:sims_gal}).
These simulations were computed at extremely high resolution: they have a force softening resolution of 60 pc, and the masses of the dark matter and gas particles are $6600 \Msun$ and $1410 \Msun$, respectively, and star particles are born at $422 \Msun$.

In contrast to the low-density environment of the {\sc Marvel}-ous Dwarfs, the {\sc D.\,C.\,Justice League} simulations are each centered on a Milky Way-analogue (table~\ref{tab:sims_gal}).
These hosts were chosen such that the suite spans a range of masses and formation histories.
This suite assumes cosmology from \citet[][]{Planck2015} ($\Omega_0$ = 0.3086, $\Omega_b$ = 0.04860, $\Lambda$ = 0.6914, $h$ = 0.67, $\sigma_8$ = 0.77;).
All four simulations are simulated at the ``Near Mint" resolution (force softening resolution of 170 pc; dark matter particle masses of $42,000 \Msun$, initial gas particle masses of $27,000\Msun$, and star particles form with masses of $8000 \Msun$).
Two of the simulations, Sandra and Elena, were also resimulated at ``Mint" resolution, close to that of the MARVEL-ous Dwarfs suite: they have a force softening resolution of 87 pc, and the masses of the dark matter and gas particles are $17,900 \Msun$ and $3310 \Msun$, respectively, and star particles are born at $994 \Msun$.
The Near Mint sample includes 108 dwarf galaxies with at least 100 particles.
The Mint sample includes the Near Mint dwarfs from the two resimulated simulations and an additional six dwarfs that fall above the resolution limit for a total of 114 dwarf galaxies.
After verifying that the resolved satellites in the Near Mint simulations available at Mint resolution had similar stellar masses and quenching times, we chose to only use the Mint versions of these two simulations. 

The {\sc Romulus25} simulation is a uniform (25 Mpc)$^3$ volume following the \citet{Planck2015} cosmology.
Its resolution is necessarily lower than the zoom-in runs.
Gas particle masses are $2.12 \times 10^5 \Msun$, about ten times higher mass than in the Near Mint simulations.
The dark matter in these runs is oversampled in order to better resolve Black Hole dynamics: $m_{DM} = 3.39 \times 10^5 \Msun$.
This lower resolution necessitates slightly different sub-grid physical recipes.
Specifically, the Romulus simulations do not include H$_{\mathrm{2}}$ physics.

\subsection{Code}
All simulations were computed using the Tree+Smoothed Particle Hydrodynamics (SPH) code, {\sc CHaNGa} \citep{Menon2015}, a descendent of the {\sc Gasoline} code \citep{Wadsley04}.
{\sc CHaNGa} uses the {\sc Charm}++ runtime system and its dynamic load balancing ensures scaling up to hundreds of thousands cores.
This modern SPH code uses a geometric mean density in the SPH force expression to prevent artificial gas surface tension \citep{Wadsley2017}, and allows for thermal and chemical diffusion across gas particles \citep{Shen10}.

The simulations track the non-equilibrium abundances of hydrogen and helium species.
In the zoom-in simulations ({\sc D.\,C.\,Justice League} and {\sc Marvel}-ous Dwarfs), these species included H$_{\mathrm{2}}$ as described in \citet{Christensen12}. 
For these calculations, H$_{\mathrm{2}}$ self-shielding and the shielding of \textsc{H\,i} by dust was calculated according to \citet{Gnedin09} using the smoothing lengths of particles for the column lengths.
Abundances of oxygen and iron and total metals were also independently tracked, and metal diffusion took place across gas particles according to a subgrid turbulent mixing modeling \citep{Shen10} with a diffusion constant of 0.03.

Heating and cooling occurs through photoionization and photoheating, collisional ionization \citep{Abel97}, $\Hmol$ collisions, radiative recombination \citep{Black81,VernerANDFerland96}, bremsstrahlung radiation, and \textsc{H\,i}, $\Hmol$ and He line cooling \citep{Cen92}.
A spatially uniform, time-dependent cosmological UV background was assumed following \citet{HaardtMadau2012}.
In the zoom-in simulations, additional H$_{\mathrm{2}}$-dissociating radiation from young stellar populations was approximated following the tree-build structure \citep{Christensen12}.
Further cooling from metal lines was calculated assuming optically thin gas in ionization equilibrium using {\sc Cloudy} \citep[version 07.02;][]{Ferland98} tables \citep{Shen10}. 

Star formation took place stochastically based on the local gas properties.
In the zoom-in simulations, these properties included the H$_{\mathrm{2}}$ abundances in order to model the observed connection between molecular clouds and star formation.
Star formation was only allowed in gas particles that were sufficiently cool ($T < 10^3$ K in the zoom-in simulations, $T < 10^4$ in the cosmological volume) and dense ($\rho > 0.1$ amu cm$^{-3}$ in the zoom-in simulations, $\rho > 0.2$ amu cm$^{-3}$ in the cosmological volume).
Note that in the zoom-in simulations the density threshold was largely superseded by the H$_{\mathrm{2}}$ dependency and the vast majority of stars form from gas with $\rho > 100$ amu cm$^{-3}$.
Each eligible gas particle of mass $m_{gas}$ had a probability of spawning a star particle of mass $m_{star}$ equal to:
\begin{equation}
    p = \frac{m_{gas}}{m_{star}} \left(1-e^{c^*\Delta t/t_{dyn}}\right)
\end{equation}
where $c_*$ is the star formation efficiency,  $\Delta_t$ is the timestep, and $t_{dyn}$ is the dynamical time.
In the zoom-in simulations, $c^* = 0.1 f_{\mathrm{H}_2} = 0.1 \frac{X_{\mathrm{H}_2}}{f_{\mathrm{H}_2} + f_{\mathrm{HI}}}$, while in the cosmological volume $c^* = 0.15$.

Each star particle represents a simple stellar population with a \citet{Kroupa2001} initial mass function.
Mass, metals, and thermal energy were returned to surrounding gas particles by Type Ia and Type II supernovae.
Stellar feedback from Type II supernovae was implemented using the ``blast-wave" feedback model \citep{Stinson06}, in which cooling was disabled for a period of time equal to the theoretical snowplow phase of the supernova \citep{McKee77}.
The zoom-in simulations assume  $\epsilon_{SN} = 1.5 \times 10^{51}$ ergs per supernova, while in the lower-resolution cosmological volume, $\epsilon_{SN} = 0.75 \times 10^{51}$ ergs per supernova.
These amounts, along with the star formation efficiencies, were tuned to reproduce (1) the stellar mass -- halo mass relation, (2) the \textsc{H\,i} gas fraction as a function of stellar mass, (3) the galaxy specific angular momentum versus stellar mass for galaxies with $10^{10.5} \Msun \leq M_{vir} \leq 10^{12} \Msun$ (notably, a mass range above where most of the galaxies in this work lie) \citep{Tremmel2017}.
The additional energy above the canonical $10^{51}$ ergs per supernova represents the energy injected into the ISM by young stars through processes such as radiation pressure.
Energy from Type I SNe was similarly distributed to nearby gas particles but without the disabling of cooling.
Mass and metals were also distributed to nearby gas particles by stellar winds, assuming the mass-loss rates from \citet{Weidemann87}.

Massive black hole formation, growth, feedback, and dynamics were modeled according to \citet{Tremmel2017} with merger criteria as in \citet{Bellovary11}.
The properties of the resulting massive black holes were analyzed in \citet{Tremmel2017} for the {\sc Romulus}25 cosmological volume and \citet{Bellovary2018} for the {\sc Marvel}-ous Dwarfs and {\sc D.\,C.\,Justice League} zoom-in volumes. A brief summary of the physics follows.
Massive black hole formation was enabled in cold, low-metallicity, overdense regions.
Star particles forming under these conditions were stochastically changed to black holes and their mass was set to $m_{BH, init}$.
Rather than ``pinning" the massive black holes to the center of their formation halos, the dynamics of the black holes were allowed to evolve naturally under the influence of the subgrid dynamical friction model from \citet{Tremmel2017}.
The massive black holes were allowed to grow via both mergers and by accretion using a modified Bondi-Hoyle prescription.
Energy from accretion was then redistributed according to the smoothing kernel. 
While massive black holes do form in some of the dwarf galaxies within the {\sc Marvel}-ous Dwarfs, the {\sc D.\,C.\,Justice League} simulations, none of them accreted at high enough rates to provide a significant source of feedback.


\begin{deluxetable}{ccc}
\tablenum{2}
\tablecaption{Description of Simulations\label{tab:sims_gal}}
\tablewidth{0pt}
\tablehead{
\colhead{Name} & 
\colhead{M$_{vir, max}$ [$\Msun$]} & \colhead{N$_{gal}$} 
}
\startdata
\multicolumn{3}{c}{{\sc Marvel}-ous dwarfs} \\
\hline
Cpt.~Marvel$^{1, 2}$& $1.55 \times 10^{10}$
& 12  \\
Rogue$^{1, 2}$     & $8.15 \times 10^{10}$
& 15 \\
Elektra$^{1, 2}$   & $4.18 \times 10^{10}$
&  14  \\
Storm$^{1, 2}$     & $7.35 \times 10^{10}$
&  27 \\ 
\hline
\multicolumn{3}{c}{{\sc D.\,C.\,Justice League}} \\
\hline
Sandra, Mint$^{3}$ & $2.4 \times 10^{11}$
&  60 \\
Elena, Mint$^{3}$& $7.5 \times 10^{11}$
&  11 \\
Ruth$^{1, 4}$      & $1.05 \times 10^{12}$
&  23 \\
Sonia$^{1, 4}$     & $1.03 \times 10^{12}$
&  20 \\
\hline
\multicolumn{3}{c}{{\sc Romulus}} \\
\hline           
{\sc Romulus}25$^5$ &  $2.29 \times 10^{13}$    
& $3273$  \\
\enddata
\tablecomments{
Simulations are included in 1: \citet{Bellovary2018}; 2: \citet{Munshi2021}; 3: \citet{Applebaum2021}; 4: \citet{Akins2021}; 5: \citet{Tremmel2017}.
}
\end{deluxetable}

\subsection{Post-processing Analysis} \label{sec:postprocess}
Individual halos were selected using {\sc Amiga's Halo Finder} \citep[AHF,][]{Knollmann2009}.
AHF identified halos by iteratively searching for over-densities using a redshift-dependent criterion.
Halo ownership was then assigned to gravitationally bound particles.
We defined the virial radius ($R_{vir}$) as the region within which the enclosed density drops below 200 times the critical density.
We also used AHF to identify satellite and central galaxies. For galaxies that lie within the same isodensity contour in a particular snapshot, the most massive halo is considered the ``central" galaxy. ``Satellites" of central galaxies are those that lie within the same isodensity contour and that overlap with the central galaxy; specifically, the distance between the halo centers must be less than the sum of the radius of the central galaxy and half the radius of the other galaxy. Backsplash galaxies are then defined to be any galaxy that is classified as a central galaxy at $z = 0$ but was classified by AHF as a satellite galaxy in a previous snapshot.

We created merger trees for each halo in order to determine the maximum halo mass and to measure the rates of mass accretion.
We used the database-generating software {\sc tangos} \citep{Pontzen2018} to identify all progenitor halos.
The main progenitor was then defined to be the most massive halo in each previous step.
Additional post-processing analysis was completed using {\sc pynbody} \citep{Pontzen2013}.

\section{Results}\label{sec:results}



\begin{figure*}
\begin{center}
\includegraphics[width=\textwidth]{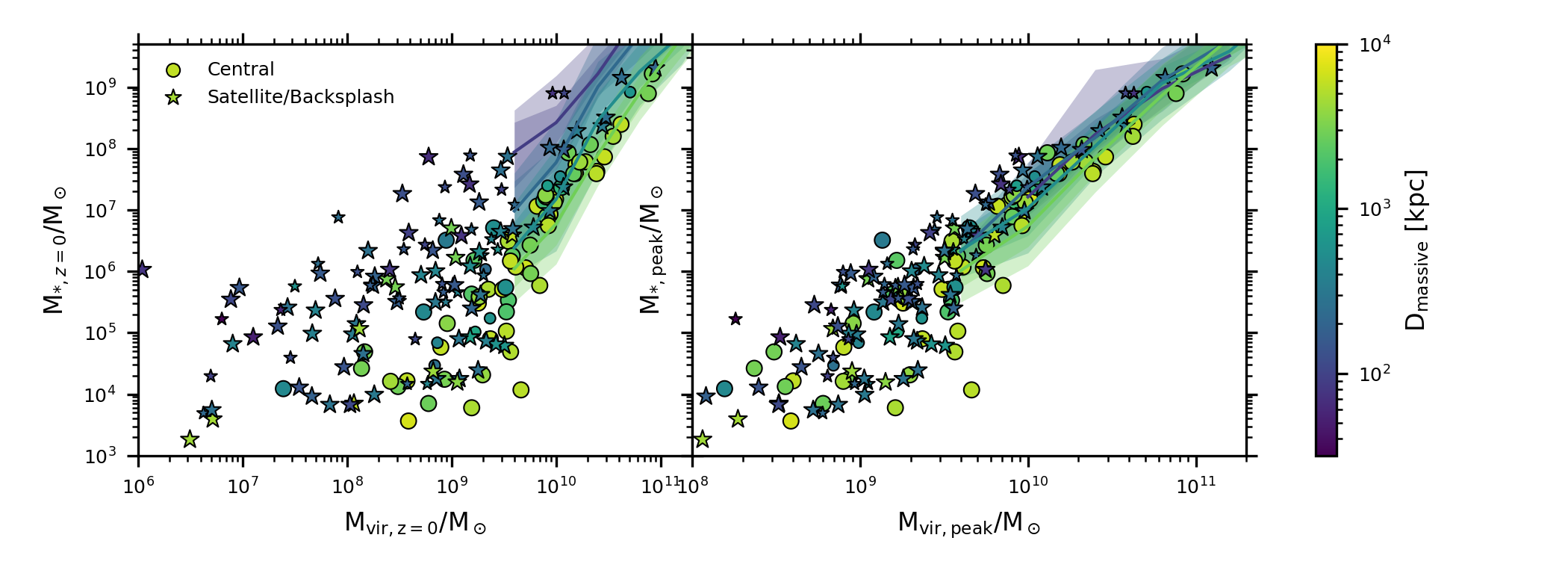}
\end{center}
\caption
{ 
The stellar mass -- halo mass relation at $z = 0$ (left) and at the time that the halo reached its maximum mass, $M_{\mathrm{vir, peak}}$ (right).
The colors show the $z = 0$ distance between the halo and a massive ($M_{\mathrm{vir}} > 10^{11.5} \Msun$) galaxy.
Data from the {\sc Marvel}-ous Dwarfs and {{\sc D.\,C.\,Justice League}} simulations are shown as points with the large points representing data from the higher resolution ({\sc Marvel}-ous Dwarfs and mint-resolution {\sc D.\,C.\,Justice League} simulations).
Stars show satellite and backsplash galaxies (for any mass host) while circles show isolated galaxies.
The four lines show the median values for the binned data from the {\sc Romulus25} simulation, with the fill regions showing the 10-90th percentile and 25-75th percentile regions.
Each line and associated regions correspond to the subset of {\sc Romulus25} galaxies in a particular distance bin; colors of these lines and regions indicate the average distance for the subset of galaxies according to the colorbar on the right.
Despite the different resolutions, cosmology, and sub-grid prescriptions, all simulations show consistent results.
The SMHM relation at $ z= 0$ shows a strong trend with environment because of the tidal stripping of the dark matter halos of satellites. 
When this effect is corrected for, as in the right panel, the trend with environment is reduced but still evident.}
\label{fig:SMHM_rom}
\end{figure*}

\subsection{Stellar Mass-Halo Mass Relation}\label{sec:SMHM}

We begin our analysis of the effect of environment on galaxy evolution with the SMHM relation.
Figure~\ref{fig:SMHM_rom} shows that relation for $z = 0$ stellar mass versus $z = 0$ virial mass for the {\sc Marvel}-ous Dwarfs, {\sc D.\,C.\,Justice League} (mint resolution when available, otherwise near-mint resolution), and {\sc Romulus25} simulations.
Satellite and backsplash galaxies (of any host mass) from the zoom-in runs are distinguished from central galaxies with the star-shaped marker.
To further illuminate the effect of environment, points are colored by the $z = 0$ distance to a massive halo (M$_\mathrm{vir} > 10^{11.5} \Msun$).
Because of the large number of galaxies in the {\sc Romulus25} cosmological volume, lines show the median values of the binned data; fill regions show the 10-90th percentile and 25-75th percentile regions.
Different colors represent the $z=0$ distances from a massive ($M_{\mathrm{vir}} > 10^{11.5} \Msun$) galaxy.
Interactions between satellites and host galaxies tend to result in a reduction of satellite virial mass to 0.005--0.5 the infall mass because of tidal stripping.
In contrast, stellar mass loss from tidal stripping is much less common with the vast majority of satellites maintaining or even increasing their stellar mass post infall.
As a result, such galaxies tend to be shifted leftward on the diagram, as shown in this graph and in other work \citep[e.g.][]{Sawala2015, Buck2019, Tremmel2020, Engler2020, Munshi2021}. 

In order to control for the effect of tidal stripping, the right-hand panel of Figure~\ref{fig:SMHM_rom} shows the stellar mass at the time of peak halo mass as a function of the peak halo mass.
While the elimination of tidal stripping is immediately apparent, the effect of environment on the SMHM relation is still evident, with galaxies closer to a more massive halo lying above and/or to the left of more distant galaxies.
It is noteworthy that despite the lower resolution, galaxies from the {\sc Romulus25} cosmological volume show the same trend as galaxies from the zoom-in simulations.
This correspondence between simulations demonstrates that {\em 1) this trend is robust across this resolution range and differences in physical parameters and 2) that the differences between the choice of cosmology for the {\sc D.\,C.\,Justice League} and {\sc Marvel}-ous Dwarfs simulations is not responsible for it.}
See Appendix A from \citet{Munshi2021} for further analysis of the (lack of) effect by the differing cosmology on the SMHM relation.
Throughout the remainder of this paper we will focus only on the zoom-in simulations to ensure that more detailed comparisons of star formation histories and ISM properties are completed for similar resolution simulations with similar sub-grid physics.

\begin{figure}
\begin{center}
\includegraphics[width=0.5\textwidth]{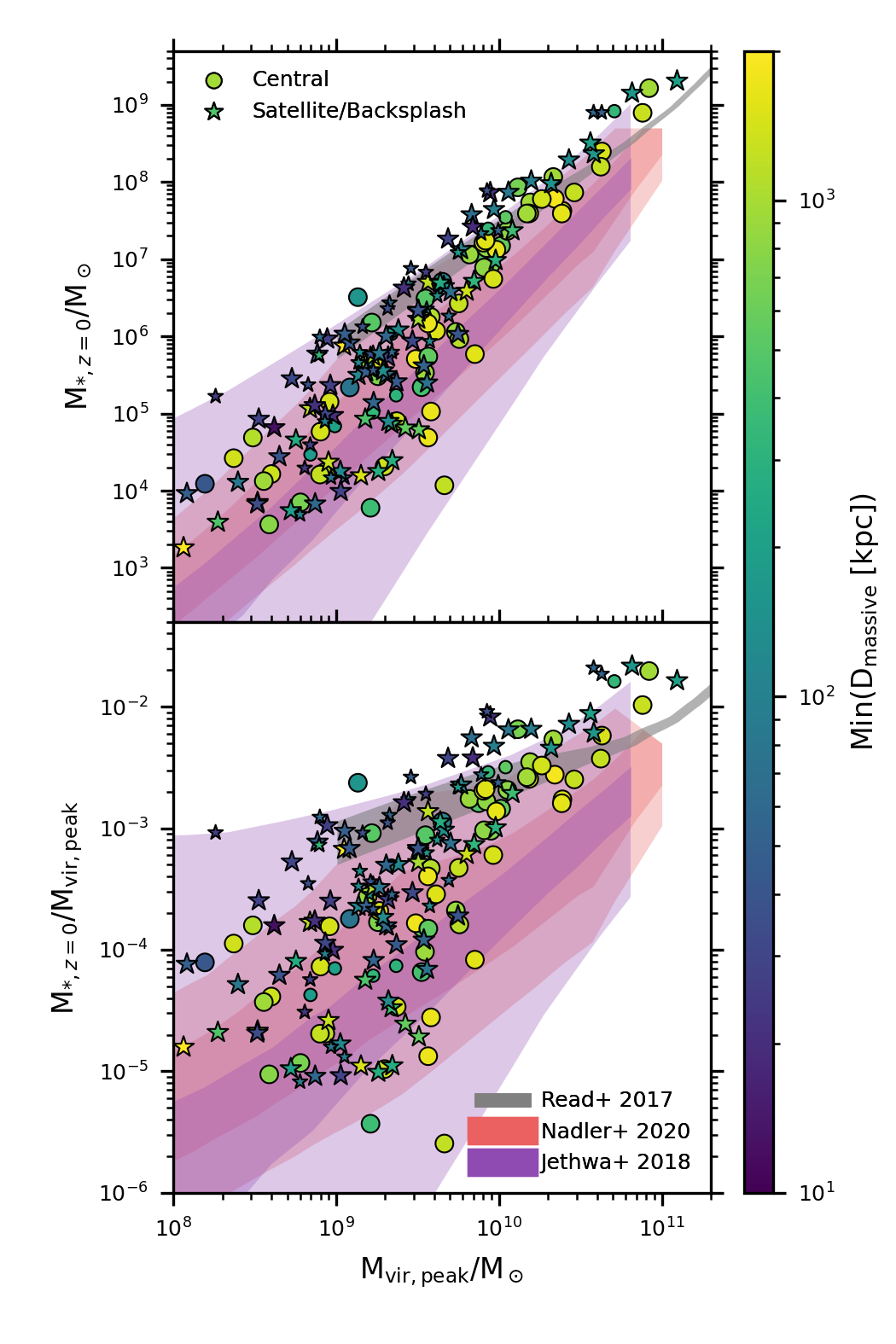}
\end{center}
\caption
{ 
The SMHM relation for the galaxies in the {\sc MARVELous} Dwarfs and {\sc D.\,C.\,Justice League} samples.
The top panel shows the $z = 0$ stellar mass versus the maximum virial mass reached by the galaxy $M_{\mathrm{vir, peak}}$, while the bottom panel shows the ratio between $z = 0$ stellar mass and $M_{vir, peak}$ as a function of $M_{\mathrm{vir, peak}}$.
Individual points are colored according to the galaxy's {\em minimum} distance to a nearest massive neighbor ($M_{vir} > 10^{11.5} \Msun$). 
Stars designate all satellites and backsplash galaxies, even galaxies that are satellites of halos less massive than $10^{11.5} \Msun$, while circles designate isolated galaxies.
The size of the points corresponds to the resolution of the simulation, as in Figure~\ref{fig:SMHM_rom}.
The colored fill regions show constraints from abundance matching: gray showing the 68\% confidence intervals from \citet{Read2017},  pink showing 68/95 \% confidence intervals from \citet{Nadler2020}, violet showing the 68/95 \% credibility intervals for $P(M_* | M_{vir})$ from \citet{Jethwa2018}.
The gradient in color of the points indicates an environmental difference with more isolated halos hosting galaxies of smaller stellar masses at fixed $M_{vir, peak}$.
}
\label{fig:SMHM}
\end{figure}

We embark on a more detailed investigation of environmental impact on the SMHM relation by showing our data in comparison to estimates for the SMHM from abundance matching (Figure~\ref{fig:SMHM}).
Specifically, we color the SMHM relation by the closest approach to the nearest galaxy more massive that $10^{11.5} \Msun$, rather than the $z = 0$ distance.
This choice of distances was made to ensure a fair comparison across halos in highly elliptical orbits whose instantaneous distance to a massive galaxy is highly variable in time.
In these plots the maximum virial mass reached by the galaxies during their lifetime, $M_{vir, peak}$, and the $z = 0$ stellar mass are used.
As in the right-hand panel of Figure~\ref{fig:SMHM_rom}, $M_{vir, peak}$ was chosen in order to reduce the effect of tidal disruption on satellite virial mass on the trend.
Unlike the right-hand panel of Figure~\ref{fig:SMHM_rom}, M$_{\mathrm{*, z = 0}}$ (not M$_{\mathrm{*, peak}}$) was used for the $y$-axis.
Our rational is that M$_{\mathrm{*, z = 0}}$ accounts for any remaining star formation that takes place after the peak halo mass is reached.
Stellar mass rarely decreases in surviving satellites following their accretion. Therefore, this measurement provides a more physically interesting comparison than what would otherwise be akin to sampling the SMHM relation at the time of peak halo mass.
It also allows us to include the six galaxies that had zero stellar mass at the time of peak halo mass in the calculations.
We note, however, that there is generally very little difference between M$_{\mathrm{*, peak}}$ and M$_{\mathrm{*, z = 0}}$, and our results are similar for either definition. 
For comparison, we show results from abundances matching studies: \citet{Read2017}, \citet{Jethwa2018}, and \citet{Nadler2020}. The data from our simulations have a scaling that is most consistent with the results from \citet{Read2017}, while having slightly elevated stellar masses compared to \citet{Jethwa2018}, and \citet{Nadler2020}. However, the scatter in our data is most similar to the results from \citet{Jethwa2018}, . While not a perfect match to any of the abundance matching studies, the level of agreement between our data and them is similar to the level of agreement between them. Further analysis of the SMHM relation for these simulations compared to abundance matching studies is provided in \citet{Munshi2021}.

We find that isolated dwarf galaxies have systematically smaller stellar masses for a given halo mass than dwarf galaxies in richer environments.
Surprisingly, the trend toward higher M$_*$/M$_{\mathrm{vir, peak}}$ with denser environments is the opposite what we would expect if interactions with the more massive galaxy simply prematurely truncated star formation. 
{\em Therefore, some other difference in the environment of satellite galaxies must play a counterbalancing role.}
For example, as will be discussed later, the environment could affect the peak halo mass reached or change the rate a which a galaxy undergoes gravitational collapse.

In order to determine the statistical significance of this environmental dependency in the SMHM relation, we compute an extra sum of squares F-test.
To do this, we first fit a single variable linear regression model to $\log$(M$_{\mathrm{*, z = 0}}$) versus $\log$(M$_{\mathrm{vir, peak}})$.
We then constructed a multiple linear regression model that included a possible dependency on the logarithm of the minimum distance between galaxy and massive neighbor, $\log$(Min($D_{\mathrm{massive}}$)), in addition to the dependency on $\log$(M$_{\mathrm{vir, peak}})$.
We found a best fit model of 
\begin{multline}
\log(M_{\mathrm{*, z = 0}}) = -13.09665 + 2.11720 \log(M_{\mathrm{vir, peak}})\\ +  -0.40125 \log(\mathrm{min}(D_{\mathrm{massive}})) .
\end{multline}
Allowing for a linear dependency of $\log$(M$_{\mathrm{*, z = 0}}$) on $\log$(min($D_{\mathrm{massive}}$)) improved our model with a significance of $2.39 \times 10^{-7}$. 
From this result, we conclude that the environmental dependency of the SMHM is significant with a very high level of confidence.

\begin{figure*}
\begin{center}
\includegraphics[width=\textwidth]{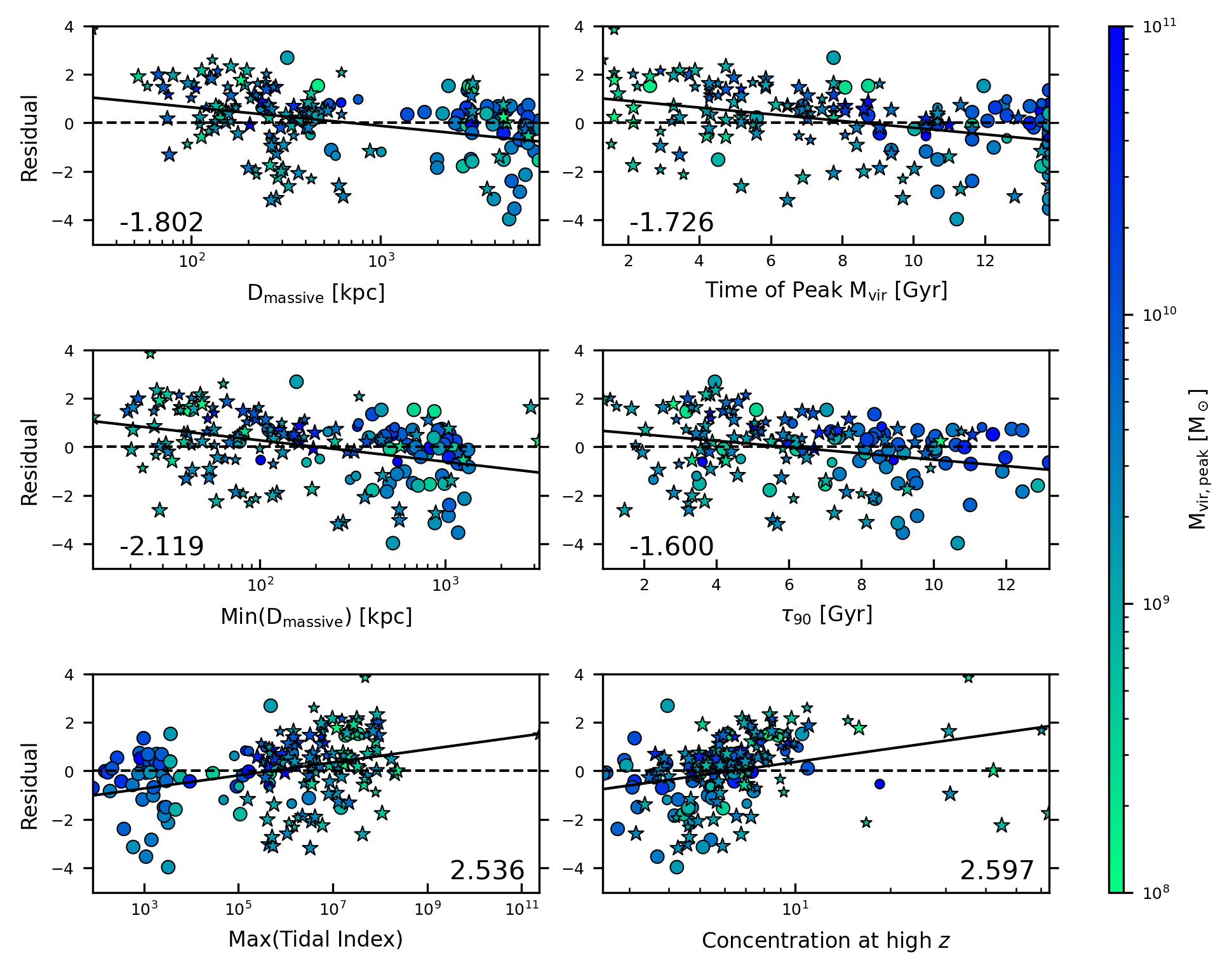}
\end{center}
\caption
{ 
The residuals from the single variable linear regression model fit to the SMHM relation are shown as a function of different possible third parameters. From top to bottom and then left to right, these parameters are the 1) z = 0 distance to a massive galaxy, 2) minimum distance to a massive galaxy, 3) maximum tidal index experienced from a perturber more massive than the target galaxy ($\propto M/r^3$ where $M$ is the mass of the perturber and $r$ is the distance to it), 4) the time at which the peak virial mass was reached, 5) the time within which 90\% of the stellar mass was formed, and 6) the concentration at high-$z$ (at the time closest to $t = 2.6$ Gyr that the galaxy is first identified, see the text for details). 
The points are colored according to the peak virial mass; stars denote backsplash and satellite galaxies and circles isolated galaxies.
The size of the points corresponds to the resolution of the simulation, as in Figure~\ref{fig:SMHM_rom}.
The dashed line marks the zero-residual line. 
The solid line shows a linear fit to the residuals versus either the log of the parameter (panels 1-3, and panel 6) or the parameter itself.
Included as text on each panel is the change in the fit line across the range of the data.
All choices of parameters show similar behavior, implying that the SMHM relation depends on them and, likely that they are correlated with each other. 
}
\label{fig:resid_all}
\end{figure*}

The potential dependency of the SMHM relation on a variety of environmental indicators can be visualized by showing the residuals from the single variable linear regression model fit versus each of those indicators.
Figure~\ref{fig:resid_all} shows such residuals  (difference between the log stellar masses from the single variable linear regression model and the log of the actual stellar masses) against the 1) z = 0 distance to a massive galaxy, 2) minimum distance to a massive galaxy, 3) square root of the maximum tidal index experienced from a perturber more massive than the target galaxy ($\propto M/r^3$ where $M$ is the mass of the perturber and $r$ is the distance to it), 4) the time at which the peak virial mass was reached, 5) the time within which 90\% of the stellar mass was formed, and 6) the concentration at high-redshift. 
If the SMHM relation has no dependency on the $x$-axis parameter, the points should appear randomly scattered around the zero-line.
However, in all cases a trend is evident in the residual and highlighted to the reader by the solid black line indicating the best-fit.
The numbers in the bottom corners of the panels are the differences in $y$-values of the best-fit lines across the range of galaxy properties, i.e. $f(x_{\textrm{max}}) - f(x_{\textrm{min}})$ where $f(x)$ is the best-fit function and $x_{\textrm{max}}$ and $x_{\textrm{min}}$ are the maximum and minimum values from the data set of the environmental indicator plotted on the $x$-axis, respectively. The motivation for providing this value, rather than the slope of the best-fit line is to better normalize across the wide range of numerical values for the different environmental indicators.

The three parameters shown in the left-hand column (D$_{\mathrm{massive}}$, Min(D$_{\mathrm{massive}}$), and Max(Tidal Index)$^{1/2}$ are all directly related to the environment of the galaxy with the square root of the tidal index inversely related to the distance to a companion.
It is, therefore, unsurprising that the residuals plotted against these parameters show similar trends, and this is further indication that the SMHM relation has a dependency on the environment.

\begin{figure*}
\begin{center}
\includegraphics[width=\textwidth]{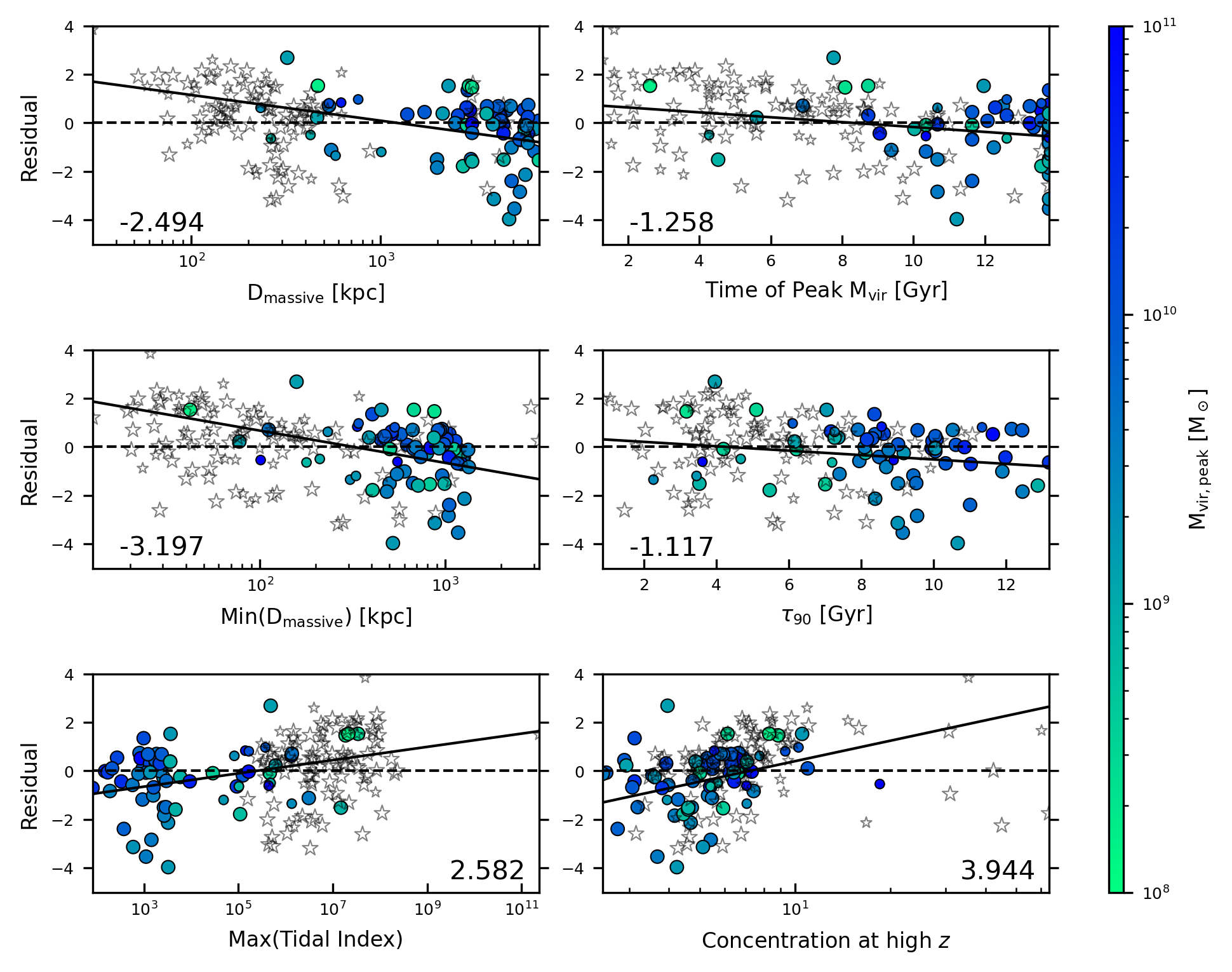}
\end{center}
\caption
{ 
The residuals from the single variable linear regression model fit to the SMHM relation are shown as a function of different possible third parameters.
Points are as in Figure~\ref{fig:resid_all}, except that here isolated galaxies are highlighted with color fill and all other galaxies are shown as open symbols.
In this plot, the solid fit line is fit only to the isolated galaxies.
Trends that appear in only the isolated galaxies are likely not the result of interactions with a more massive galaxy.
}
\label{fig:resid_partial}
\end{figure*}

The right-hand three panels are all related to the timescale of galaxy formation.
Galaxies that grow faster (or have their growth truncated sooner) will reach their peak virial mass in less time, form 90\% of their stars in less time, and may exhibit higher concentrations \citep[e.g.,][]{Wechsler2002, Zhao2003}.
Comparing concentrations across galaxies raises some challenges, as concentrations (defined as the scale radius/$R_{\mathrm{vir}}$) will decrease over time as the virial radius grows, even in the absence of gravitational interactions with a larger host.
Therefore, examining the concentration at the time of peak halo virial mass would result in lower concentrations for galaxies that reach their peak mass later, even in the absence of any other environmental dependency.
Instead, we compare the concentrations of the galaxies at $t = 2.6$ Gyr (or the closest time after that, if the galaxy progenitor was not yet identifiable at that snapshot).
This time was chosen to be early enough that most galaxies (84 out of 95) had not yet been accreted by a host but late enough that most galaxies (83 out of 95) already had an identifiable progenitor.
We further note that the Planck cosmology used in the {\sc D.\,C.\,Justice League} simulations results in a denser high-redshift Universe than the {\sc Marvel}-ous Dwarfs simulations, we could result in higher concentrations. 
Happily, though, the trend in the residuals with concentration is similar and even stronger if only the isolated galaxies (which are almost entirely from the {\sc Marvel}-ous Dwarfs simulations) are included (see figure~\ref{fig:resid_partial}), so the trend does not appear to be due to the different cosmologies used.
We find that galaxies that lie above the SMHM relation formed in shorter amounts of time (i.e., had lower values of ``Time of Peak M$_{\mathrm{vir}}$'' and ``$\tau_{90}$'') and had higher concentration at early times.
One possible explanation for this dependency that we discuss further in \S\ref{sec:mass_assem} is that the position of these galaxies above the SMHM relation results from an underlying evolution of the SMHM with time.

One remarkable feature of the environmental dependency of the SMHM relation is that it appears even when satellite and backsplash galaxies are not included in the analysis.
Figure~\ref{fig:resid_partial} highlights this feature by showing the trends in the residuals that arise when only the isolated galaxies are considered.
From this plot, it is evident that more isolated central galaxies and central galaxies that assemble later tend to lie below the SMHM, in contrast to less isolated and more rapidly forming central galaxies.
Moreover, the change in fit lines across the entire range of each parameter (shown as the numbers in each panel of Figure~\ref{fig:resid_all} and Figure~\ref{fig:resid_partial}) are similar whether or not backsplash and satellite galaxies are included.
Limiting our analysis to only isolated galaxies necessarily reduces the robustness of our results; while still statistically significant, $p$ was increased to 0.0075. 
The persistence of these trends outside of the virial radius of the more massive galaxy indicates that direct interactions with a massive galaxy cannot be solely responsible for them.
Other mechanisms by which a denser environment can affect galaxy evolution must, therefore, be considered.

\subsection{Baryonic Content}\label{sec:BMHM}
\begin{figure}
\begin{center}
\includegraphics[width=0.5\textwidth]{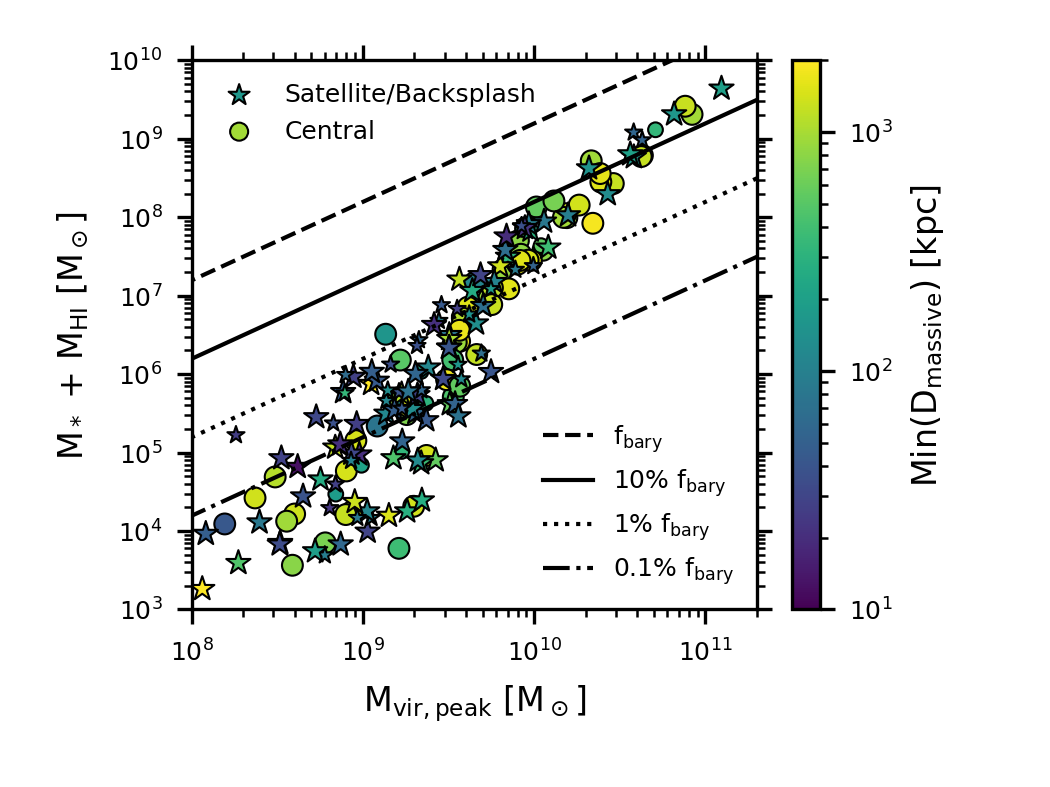}
\end{center}
\caption
{ 
The $z = 0$ baryonic mass of the galaxies versus their peak halo mass, colored according to their minimum distance to the nearest massive halo ($M_{\mathrm{vir}} > 10^{11.5} \Msun$). 
Stars designate all satellites and backsplash galaxies while circles designate isolated galaxies.
The size of the points corresponds to the resolution of the simulation, as in figure~\ref{fig:SMHM_rom}.
The lines show various percentages of $f_{bary} \times M_{\mathrm{vir}}$, assuming a Planck cosmology. 
Unlike the stellar mass, the baryonic masses of the galaxies for a given halo mass do not appear to strongly depend on environment. 
}
\label{fig:bmhm}
\end{figure}

Larger stellar masses for a given halo mass could be the result of either a greater baryonic mass or a greater efficiency of star formation.
We, therefore, show the Baryonic Mass -- Halo Mass relation, where the baryonic mass is defined to be sum of the \textsc{H\,i} and stellar mass, i.e., the material that would be considered part of the disk (Figure~\ref{fig:bmhm}).
Unlike the SMHM relation, the Baryonic Mass -- Halo Mass relation shows little trend with environment. 
Galaxies lie along the same trend regardless of whether they are isolated or satellite galaxies or how far they are from a massive galaxy.
When calculating the improvement to a linear regression model by including a dependency on $\log(\mathrm{min}(D_{\mathrm{massive}}))$, we find a best fit model of 
\begin{multline}
\log(M_{\mathrm{*, z = 0}} + M_{\mathrm{HI}}) = -14.91675 + 2.27170 \log(M_{\mathrm{vir, peak}}) \\
+  -0.16362 \log(\mathrm{min}(D_{\mathrm{massive}}))
\end{multline}
with the significance of improvement only $p = 0.02859$ (compared to $p = 2.391\times 10^{-7}$ for the SMHM relation for the entire sample).
Not only is the significance of the result reduced, the best-fit dependency of the baryonic mass on the minimum distance is weaker than it was for the stellar mass (-0.16362 dependency compared to -0.40125).
From this lack of a strong trend, we conclude that environment plays little role in the mass of baryons within the disk at $z = 0$ and instead primarily affects the efficiency by which those baryons are transformed into stars.
This comparison indicates that isolated galaxies are, as we will demonstrate further below, slower to form stars.
Notably, if the total baryonic mass (i.e., including CGM material) at the time of peak halo is considered, we see a similar lack of trend.

The fact that the relation between baryonic mass and halo mass is tighter than for stellar mass and halo mass may not be surprising in light of the baryonic Tully-Fisher relation.  It has been known for quite some time that stellar mass as a function of velocity (where velocity is a proxy for halo mass) shows more scatter than stellar mass $+$ \textsc{H\,i} mass \citep[e.g.,][]{McGaugh2000}.  However, studies of the baryonic Tully-Fisher relation generally exclude satellites \citep[e.g.,][]{geha06, McGaugh2012, Lelli2016, McQuinn2022}.  There is no {\it a priori} reason to expect that satellite galaxies should follow a similar trend to field galaxies.  
On the contrary, the reduced trend between the baryonic mass and environment seen in Figure~\ref{fig:bmhm} is especially surprising given that ram pressure stripping is known to remove gas from satellite galaxies \citep[e.g.][]{murakami_interaction_1999}.
It appears that this gas loss, though important for satellite quenching, is a relatively small amount of the total baryonic disk mass, although it may have a larger effect if the amount of satellite halo gas is also considered.

\subsection{Star Formation Histories}\label{sec:sfh}


\begin{figure}
\begin{center}
\includegraphics[width=0.5\textwidth]{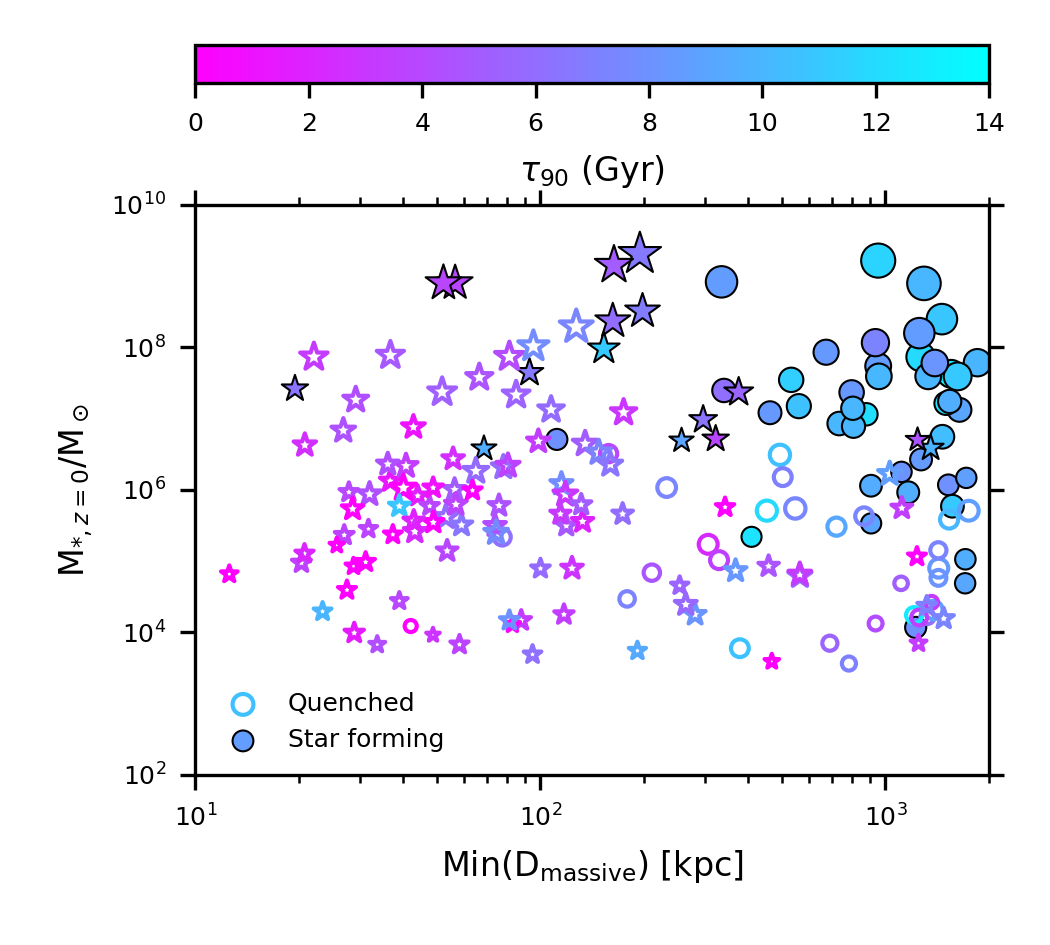}
\end{center}
\caption
{ 
The time at which 90\% of the star formation had taken place ($\tau_{90}$, shown by the color map) for galaxies of different stellar masses and distances from their nearest massive neighbor.
The size of the points are scaled according to their virial radii.
Quenched galaxies are indicated by open symbols and star forming galaxies by solid.
Empty magenta/purple points generally represent galaxies quenched by reionization.
Empty blue stars at small distances show galaxies likely quenched by satellite processes.
Isolated galaxies not only experience less quenching at all stellar masses, their star formation happens later (i.e., larger values of $\tau_{90}$) than satellite galaxies of the same stellar mass. 
}
\label{fig:smassdist}
\end{figure}

The observed trends in likelihood of quenching, color, and gas fraction for satellite vs field galaxies \citep[e.g.][]{Geha2012} should all be related to large-scale differences in the star formation histories.
Therefore, we compare $\tau_{90}$, the time at which 90\% of the star formation had taken place, as a quantitative measure of the history of star formation that is closely related to the galaxy color.
(Later, in \S\ref{sec:obs} we draw additional and more one-to-one comparisons to observations.)

Figure~\ref{fig:smassdist} shows the values of $\tau_{90}$ for quenched and star-forming galaxies of different stellar masses and degrees of isolation.
Galaxies of the same stellar mass are more likely to be quenched if they are near a massive galaxy. Some of this trend, especially within a few hundred kiloparsecs (similar to the virial radii of the halo), may be directly attributable to quenching of satellite and backsplash galaxies by the main halo.
In general, the smaller minimum distances to massive galaxies correlates with earlier star formation.

More isolated galaxies also have higher values of $\tau_{90}$, and the plot highlights a particular population of low-mass, isolated galaxies that are star forming at late times. 
 The most extreme examples of these galaxies are the few highly-isolated, star forming galaxies with $M_* < 10^6 \Msun$ visible in the lower-right portion of this figure.
 As will be further discussed when comparing the cumulative star formation histories of these galaxies, these star forming galaxies have substantially higher $M_{\mathrm{peak, vir}}$ than quenched galaxies of similar stellar masses.
 The larger halo masses of the low-stellar-mass, star forming galaxies evidently enabled them to retain or re-accrete gas following reionization.

\begin{figure*}
\begin{center}
\includegraphics[width=\textwidth]{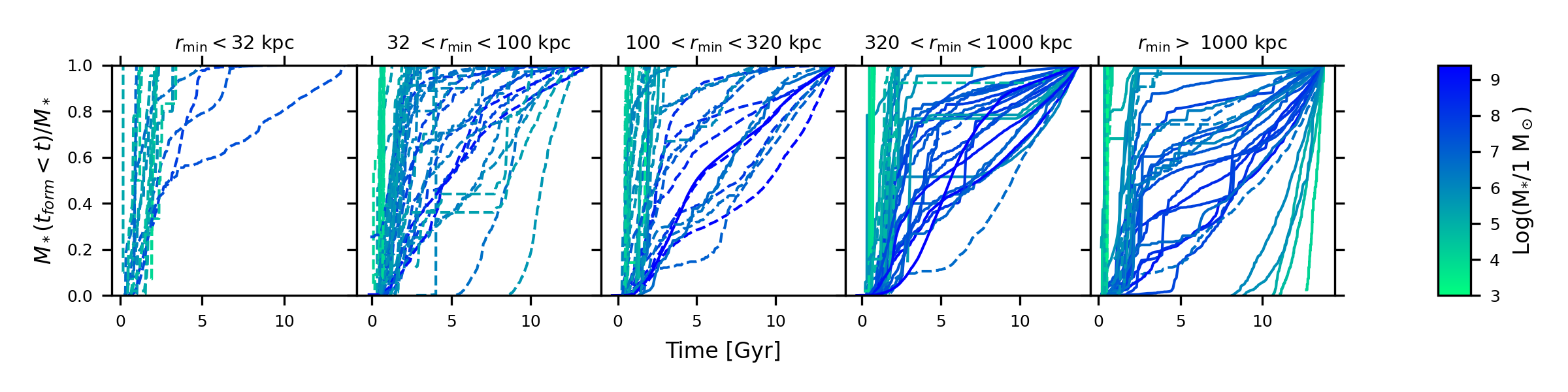}
\end{center}
\caption
{ 
Normalized, cumulative star formation histories for each of the galaxies, subdivided into different panels by their minimum distance to a massive galaxy.
Satellite and backsplash galaxies (regardless of host mass) are indicated by dashed lines while the colors represent the stellar mass .
}
\label{fig:sfh}
\end{figure*}

Figure~\ref{fig:sfh} shows the cumulative star formation histories of the individual galaxies in different minimum distance bins.
The greater likelihood of quenching for galaxies near massive hosts is clearly evident.
In general, more isolated galaxies undergo later star formation.
This result is consistent with the star formation histories of dwarf galaxies from the APOSTLE and Auriga simulations \citep{Digby2019}, NIHAO \citep{Buck2019}, and TNG50 \citep{Joshi2021}.
The latter found clear differences between the SFHs of satellite and central galaxies, and more subtle differences when comparing the SFHs of galaxies with M$_*$ between 10$^{7.5-8.0} \Msun$ at different distances from a massive galaxy (we note, though, that all of their star formation histories were more consistent with a constant star formation model prior to quenching than ours).
Our results are also consistent with \citet{Garrison-Kimmel2019}, who found that ``highly" isolated central dwarf galaxies with $M_* \leq 10^7 M_\odot$ had later star formation histories than both satellite and central galaxies within Local Group and single MW-mass host environments.
As in our simulations, some of their most isolated low-mass galaxies formed the vast majority of their stars after 9 Gyrs.
Even within a virial radius, there is an indication that satellites on the outskirts of the host halo experienced later star formation than more centrally located satellites, as also seen in \citet{Engler2022, Joshi2021}.
Presumably, these more distant satellites have experienced more recent infall and, thus, later quenching by the host.

Especially at the closest distances, the star formation histories tend to be weighted toward earlier times for galaxies of lower stellar mass.
While at all distances, the galaxies with $M_* \leq 10^6\Msun$ typically quench within the first few Gyrs, this trend between the stellar mass and shape of star formation history becomes less clean at larger distances.
The farthest distance bin, in particular, shows a variety of star formation histories, including reigniting star formation and late-starting star formation that begins around 10 Gyrs.

The lowest stellar mass galaxies at the greatest distance are most likely to show evidence of reignition of star formation.
In this model, star formation ceases early on ($\sim 4$ Gyr), likely because of reionization-driven gas loss. 
Sometime later star formation is reignited and continues until late times.
In the plots of cumulative star formation history, this pattern appears as a horizontal line with a rise on either side. 
Observationally, both Leo A \citep{Cole2007} and DDO 210 \citep{Cole2014} show signatures of this type of star formation
While not a perfect analog, the star formation histories of the three late-forming\footnote{We note that \citet{Bozek2019} suggested such late-forming dwarfs were a product of warm dark matter scenarios, as they did not find any in CDM simulations.  On the contrary, we are finding late-forming dwarfs with a CDM context.  However, more work is required to determine if they are numerically robust.} dwarf galaxies are reminiscent of KDG 215.
This gas-rich, low surface brightness galaxy reached a peak star formation rate $\sim 1$ Gyr ago \citep{Cannon2018}. 
Previous simulation work has shown that such ignitions of star formation can be triggered by encounters with gas streams within the IGM \citep{Wright2018}, by mass growth patterns in which a halo oscillates around a critical star forming mass \citep{BenitezLlambay2015, Fitts2017, PereiraWilson2022}, 
 or simply through the accumulation of gas accompanying a slowly-growing dark matter halo \citep{Rey2020}. 

Considering the role of virial mass, rather than stellar mass can explain some of the diversity in the star formation histories. 
In particular, neither the late-starting nor reigniting galaxies inhabit the smallest mass halos, instead residing in halos of $M_{\rm peak, vir} \sim 10^{9.1} - 10^{9.8}$ M$_{\odot}$.
This mass is similar to the threshold below which reionization suppresses star formation \citep[e.g.][]{Okamoto08}.
Indeed, we see that yet smaller mass halos ($M_{\rm peak, vir} \lesssim 10^9 \Msun$) at both near and far distances contain galaxies quenched within the first 5 Gyr.
These are presumably galaxies whose star formation ceased because reionization either removed gas from the halo or halted further accretion onto it.
In the late-starting and reigniting galaxies, the slightly higher peak halo mass presumably enabled periodic growth of the halo beyond the reionization threshold, which allowed for the later accretion of gas and the (re)ignition of star formation \citep[e.g., the reigniting galaxies from][]{Fitts2017}.

\subsection{Mass Assembly History}\label{sec:mass_assem}
One driver of differences in star formation histories may be differences in the mass assembly histories. 
This connection is further supported by the tendency in hierarchical formation for galaxies forming in denser environments to assemble their mass more quickly \citep[e.g.,][]{Wechsler2002, Diemand2005}. 
It may be that the differences in the star formation histories outlined in the previous section originate from environment-driven differences in the mass assembly histories.
We examine this hypothesis by comparing the timescales for both the virial and stellar mass growth.

\begin{figure*}
\begin{center}
\includegraphics[width=\textwidth]{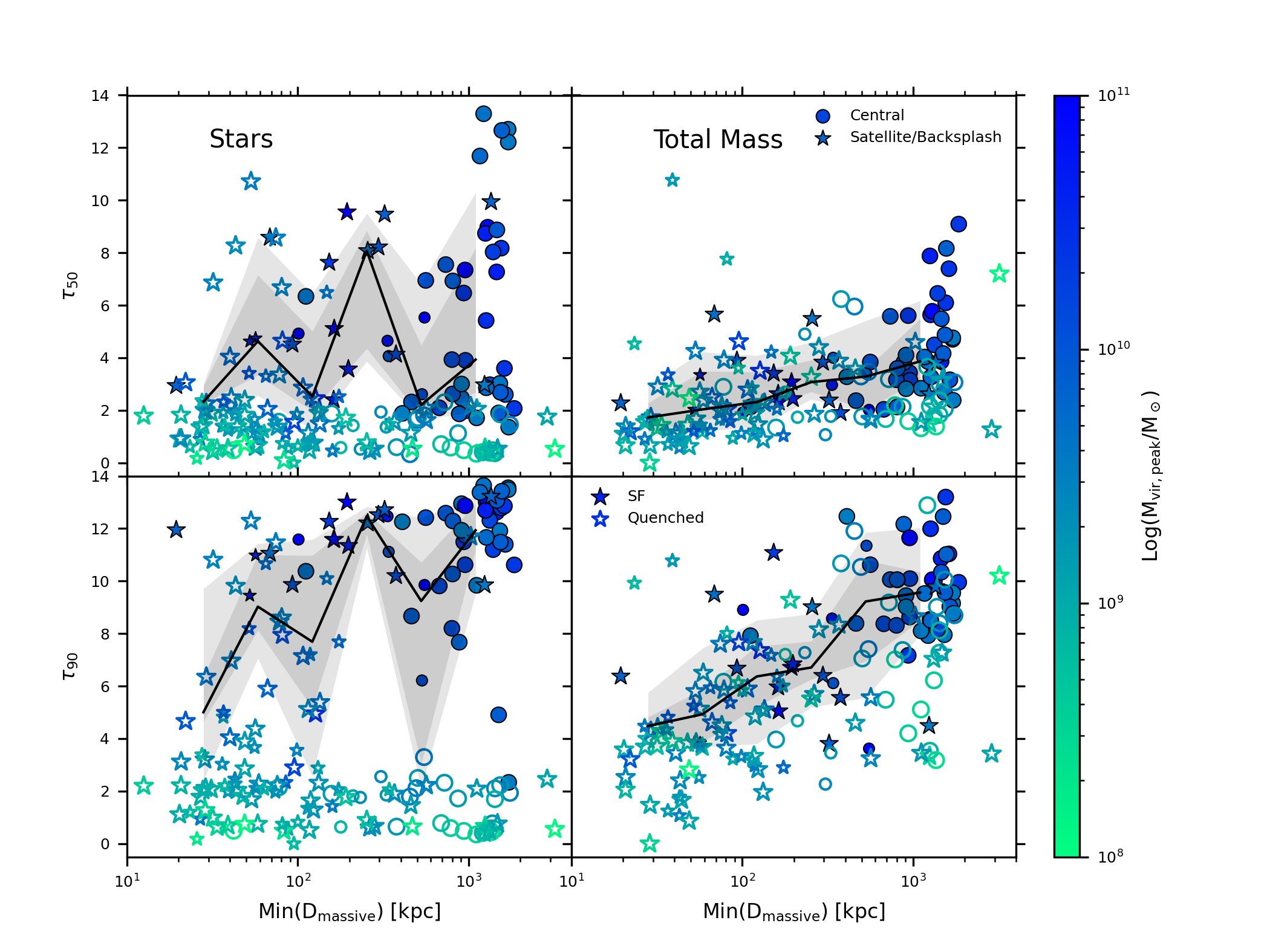}
\end{center}
\caption
{ 
Comparisons between the star formation and mass assembly histories of galaxies in different environments. 
The left panels show the assembly of the stellar mass and the right panels show the assembly of the total mass.
The top row measures the time at which 50\% of the peak mass is obtained and the bottom the time at which 90\% of the peak mass is obtained.
Filled symbols designate star forming galaxies while empty points designate quenched galaxies.
Colors indicate the peak virial mass of the galaxy.
Stars designate all satellites and backsplash galaxies while circles designate isolated galaxies.
The size of the points corresponds to the resolution of the simulation, as in Figure~\ref{fig:SMHM_rom}.
The filled area shows the 10-90th and 25-75th percentile regions for the galaxies with $M_{\mathrm{vir, peak}} > 10^{9.5} \Msun$ (the mass range unlikely to be quenched by reionzation), while the black line shows the median.
More isolated galaxies tend to assemble their stellar mass later (as seen by $\tau_{90}$) and their total mass later (as seen by both $\tau_{50}$ and $\tau_{90}$)
}
\label{fig:taus}
\end{figure*}

Figure \ref{fig:taus} shows the times at which the stellar and virial masses of the main progenitor grew to 50\% ($\tau_{50}$) and 90\% ($\tau_{90}$) of their maximum as a function of environment.
For both the stellar and the virial mass there is some trend between $\tau_{90}$ and  environment.
The trend between $\tau_{50}$ and environment is fainter, indicating that any differences in the assembly history caused by environment tend to grow over time.
These trends are shown across galaxies of all mass, but also hold when comparing galaxies of similar values of M$_{\mathrm{vir, peak}}$.
The exception is the stellar $\tau_{50}$ and $\tau_{90}$ for galaxies with M$_{\mathrm{vir, peak}} \lesssim 3 \times 10^{9} \Msun$.
For galaxies with M$_{\mathrm{vir, peak}} \lesssim 3 \times 10^{9} \Msun$, $\tau_{50}$ and $\tau_{90}$ for the {\em stellar mass} are uniformly low ($\lesssim 3$ Gyr), regardless of environment. 
These low values are a sign that the cosmic UV background halted star formation in all of these galaxies by the time of reionization.
Since for these galaxies star formation is halted, even as the galaxy continues to grow in total mass, the relationship between stellar mass and dark matter mass timescale breaks down.
Given that our simulations assume a spatially constant cosmic UV background, it is unsurprising that we do not see an environmental trend in the reionization quenching of the lowest mass galaxies.
However, in the actual Universe, patchy reionization could well alter this pattern \citep[e.g.,][]{Wu2019, Katz2020, Ocvirk2020}.

One difficulty in interpreting this figure is that the growth of both virial mass and stellar mass is halted when a galaxy becomes a satellite of a more massive galaxy.
This halting of mass growth necessarily shifts $\tau_{50}$ and $\tau_{90}$ to earlier times, independent of differences in the earlier mass assembly.
We note, though, that the trend toward later values of $\tau_{50}$ and $\tau_{90}$ exist even when examining galaxies outside of the virial radius of a massive host.

\section{Discussion}\label{sec:discussion}

\subsection{Evolution of the Stellar Mass-Halo Mass Relation}\label{sec:evol}

As shown in Figure~\ref{fig:resid_all}, galaxies that reached their peak virial mass earlier tend to lie above and to the left of the SMHM relation.
In other words, they have higher $z = 0$ stellar masses for a given $M_{\mathrm{vir, peak}}$ than galaxies that reached their peak virial mass later.
As further discussed in \S\ref{sec:mass_assem}, galaxies that reached their peak virial mass earlier also formed in denser environments, as measured by their minimum distance to a galaxy with $M_{\mathrm{vir}} > 10^{11.5} \Msun$.
One possible interpretation of these two trends is that the environmental dependency of the SMHM relation originates in the ``freezing-out" of galaxies in denser environments at earlier times.
Essentially, galaxies in denser environments assemble their mass earlier so they reach their peak halo mass and form the bulk of their stellar mass at earlier times.
Their position in $M_{\mathrm{*, z = 0}}$ versus $M_{\mathrm{vir, peak}}$ space is, therefore, a relic of the SMHM relation from that earlier time.

\begin{figure}
\begin{center}
\includegraphics[width=0.5\textwidth]{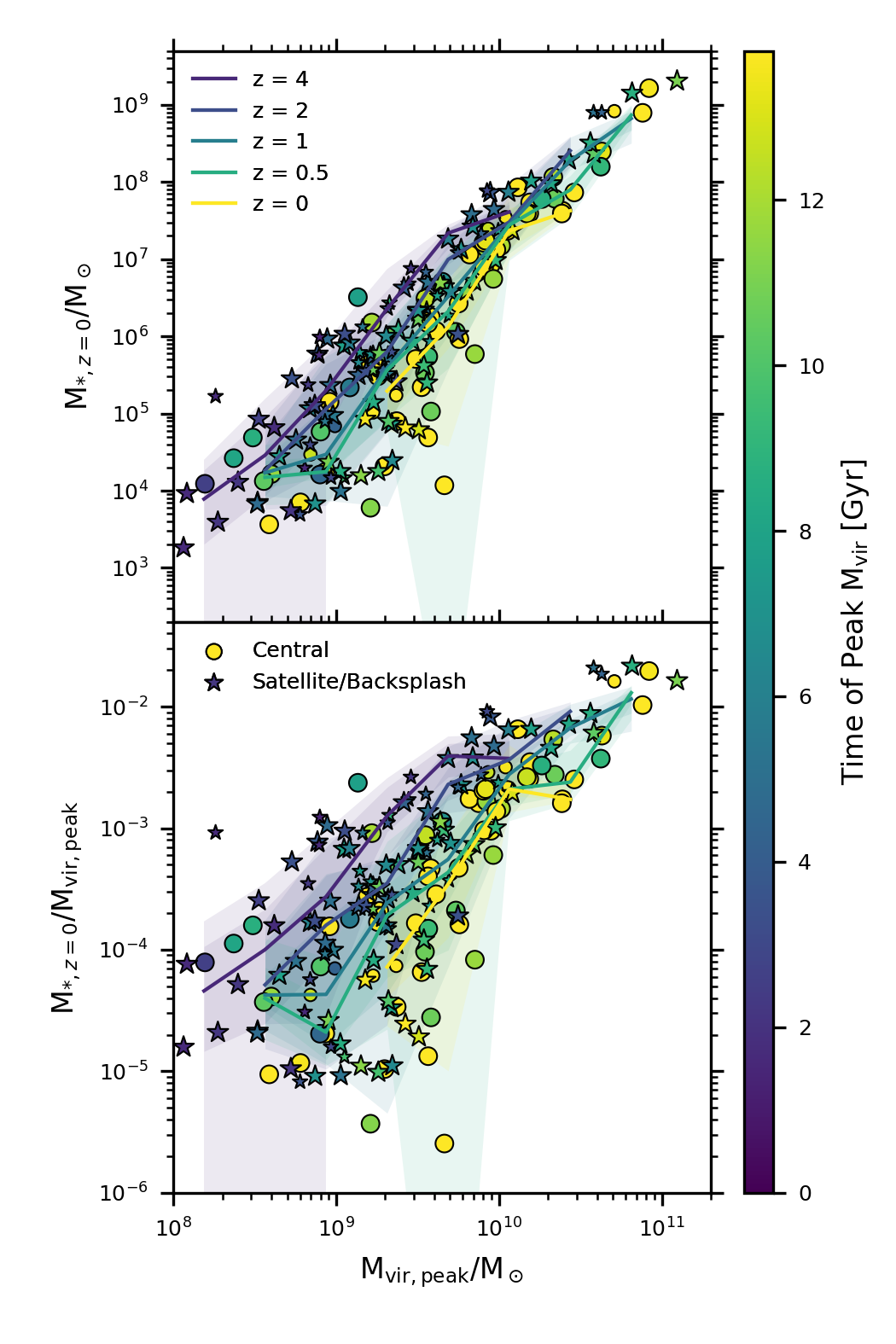}
\end{center}
\caption
{ 
The SMHM relation for the galaxies in the {\sc MARVELous} Dwarfs and {\sc D.\,C.\,Justice League} samples, colored according to the time they reach peak virial mass.
The top panel shows the $z = 0$ stellar mass versus the maximum virial mass reached by the galaxy $M_{\mathrm{vir, peak}}$, while the bottom panel shows the ratio between the $z = 0$ stellar mass and $M_{\mathrm{vir, peak}}$ as a function of $M_{\mathrm{vir, peak}}$.
Individual points are colored according to the time at which the galaxy reaches its peak virial mass.
As in Figure~\ref{fig:SMHM_rom}, stars designate all satellites and backsplash galaxies, even galaxies that are satellites of halos less massive than $10^{11.5} \Msun$, while circles designate isolated galaxies; the size of the points corresponds to the resolution of the simulation.
The solid lines show the median SMHM at different redshifts ($z = 3, 2, 1$, $0.5$, and $0$) for the subset of galaxies that have not yet reached their peak virial masses at that time.
These galaxies are binned by $M_{\mathrm{vir, z}}$, the virial mass at that redshift, and the corresponding transparent color-fill regions designate the 25$^{\mathrm{th}}$--75$^{\mathrm{th}}$ and 10$^{\mathrm{th}}$--90$^{\mathrm{th}}$ percentile regions.
The correspondence between the lines and points of similar colors illustrates that the location of a galaxy in $M_{\mathrm{*, z = 0}}$ versus $M_{\mathrm{vir, peak}}$ space is consistent with the underlying evolution of the SMHM relation.
}
\label{fig:SMHMevol}
\end{figure}

We examine this interpretation by comparing $M_{\mathrm{*, z = 0}}$ versus $M_{\mathrm{vir, peak}}$ for galaxies that reach their peak virial mass at different times to the underlying SMHM at different redshifts.
Figure~\ref{fig:SMHMevol} shows $M_{\mathrm{*, z = 0}}$ versus $M_{\mathrm{vir, peak}}$ colored according to the time of peak $M_{\mathrm{vir}}$ on top of the median SMHM relation at different redshifts for only the subset of galaxies that had not yet reached peak $M_{\mathrm{vir}}$ at that redshift.
In other words, the median SMHM relation at higher redshifts was determined for only those galaxies that were still growing their virial mass.
This figure shows a telling correspondence between the shape of the SMHM relation at different redshifts and the points of similar colors.
In these simulations, the SMHM relation moves leftward over time, indicating that the ratio of stellar mass to virial mass is reduced at lower redshfits.
This shift is consistent with the tendency of galaxies in denser environments to lie 
above the median $z = 0$ SMHM relation, implying that their ratio of stellar mass to virial mass may reflect the properties of higher-redshift galaxies and a cessation of their stellar growth since that earlier time.

One straightforward explanation of this ``freezing-out" of galaxies in denser environments at earlier times may be described through satellite-host interactions.
In this scenario, the accretion of a satellite galaxy onto a more massive host both causes the virial mass to decrease through tidal stripping while roughly simultaneously quenching star formation.
Therefore, the $M_{\mathrm{*, z = 0}}$ and $M_{\mathrm{vir, peak}}$ of the satellite are approximately that of the satellite at infall.
We caution, however, that this highly-plausible scenario for satellites of massive hosts does not explain the similar trends we see for galaxies at intermediate distances.
Even in the absence of direct interactions with a host, these galaxies reach their peak virial mass and form the bulk of their stars before $z = 0$.
Such galaxies may undergo a reduction in their virial mass over time, either through passing interactions with other low-mass galaxies or if it enters into a higher-density region.
In the latter case, the higher ambient density may result in $R_{vir}$ being defined at a smaller distance, with less mass enclosed.
Regardless, it is likely that the freezing-out at different times of halos in different environments is a universal phenomenon that may or may not include direct host-satellite interactions for a given galaxy.
Finally, if this freezing out of halos at different redshifts is responsible for the environmental dependency we discuss here, it may be a distinguishing feature across simulations and useful for testing different physical models.

Additionally, we note that the very concept of the SMHM relation is based on the definition of halo mass.
Previous works have shown that M$_{\rm star}$ correlates more strongly with v$_{\rm peak}$, the peak maximum circular velocity, due to the effects of halo assembly bias \citep[e.g.,][]{ChavesMontero2016, Reddick2013}. 
Since $M_{vir}$ is defined to be the mass enclosed within a specific density threshold (in this case, 200 times the critical density at that redshift), changes to the underlying density can affect the value of M$_{vir}$. 
For example, subhalos that reside within a higher-density region could have lower measured values of  M$_{vir}$ than the same galaxy formed within a lower-density region.
Other measurements of halo mass can produce different results. 

In particular, we find no environmental dependency for the stellar mass when the peak maximum circular velocity is considered rather than the peak virial mass.
However, then isolated galaxies contain higher amounts of disk baryons (\textsc{H\,i} + $M_{*}$) than less-isolated galaxies of the same peak maximum circular velocity.
In other words, an environmental dependency for the stellar mass is replaced by an environmental dependency for the disk baryon mass when maximum circular velocity is considered rather than virial mass.
Since $M_{vir}$ remains the more widely used measurement of mass and since it does not depend on the entire mass distribution of the galaxy, we have chosen to focus our analysis on it.
{\em Regardless of how the halo mass is defined, though, the trends in star formation (and the associated quantities, such as gas fraction and color) with environment remain.} We discuss those observational implications below in Section \ref{sec:obs}.

\subsection{Comparison with other theoretical work}\label{sec:theory}
Previous theoretical work has also examined the effect of environment on the SMHM relation with mixed results.
The trend toward higher $M_*/M_{\mathrm{vir, peak}}$ and environment is consistent with the data presented in \citet{Sawala2012}.
Their analysis was completed on the AQ-C-5 Milky Way-analogue simulation from the `Aquarius’ project' \citep{Springel08}, which has a mass resolution a couple of orders of magnitudes lower than our highest resolution simulations.
Their simulations only include dwarf galaxies formed within the resolved region proximate to the Milky Way analog. 
Nevertheless, they find a slight trend toward higher $M_*/M_{\mathrm{vir, peak}}$ values for satellite galaxies than non-satellites.
They attribute this trend in their data to the tendency for satellites to reach their peak halo mass earlier and to their halo identification algorithm requiring a higher density when the mean background density is higher, as is the case for satellites.
Their first explanation is entirely consistent with our analysis showing that the environmental dependency of the SMHM relation to be consistent with different galaxies ``freezing-out" of an evolving SMHM relation.

This trend between $M_*/M_{\mathrm{vir, peak}}$ in denser environments is also consistent with \citet{Arora2022} who found that when comparing across $z = 0$ (rather than peak) virial mass, simulated central dwarf galaxies from a Local Group environment had stellar masses that were 0.2-0.3 dex higher than those with the same virial mass from an isolated simulation.
Since they only consider central galaxies, one would expect minimal virial mass loss from tidal stripping; however, it is worth noting that they compared dwarf galaxies of similar M$_{\mathrm{vir, z = 0}}$, rather than M$_{\mathrm{vir, peak}}$.
They conclude that this difference likely stems from the pre-enrichment of gas from other galaxies, which could result in more rapid gas cooling and earlier star formation.
We note that their results differ from ours, though, in that they find increased amounts of cold gas in dwarfs within Local Group-like environments.

In contrast, \citet{Shi2020}, found no difference in the SMHM relation between satellites and host galaxies in the IllustrisTNG 100 Mpc cosmological volume. 
However, their work only extended down to peak virial masses of $4 \times 10^{10} \Msun$.
It is not clear that we find a difference in the SMHM relation in this mass range, as we have few data points for these massive of galaxies and the greatest differences we observe are for lower masses.
\citet{Buck2019} also finds no environmental difference in the SMHM for their simulations. 
However, their simulations only allow them to examine distances out to 2000 kpc (almost a factor of four smaller than our analysis), and it is not clear from the data that a more extensive statistical analysis over a longer distance range would produce different results than ours.

While not examining the SMHM relation directly, \citet{Joshi2021} compared the stellar assembly times for low-mass galaxies ($10^7$--$10^{10}$ M$_*$) across a range of environments using the TNG50 cosmological volume. 
In their analysis, they compared the star formation histories for satellites up until to time of accretion to a control sample of isolated dwarf galaxies that were of similar virial mass at the same time.
They found some evidence for earlier assembly of the stellar mass for the satellites, especially when examining satellites of the most massive hosts ($M_{\mathrm{vr}} 10^{14.0-14.3}$).
For this subsample, the satellites had formed $\sim$15-30 percent more of their stellar mass at the time of accretion.
They are able to attribute all of this difference to preprocessing of the satellites by lower mass hosts.
However, unlike our results, they see no evidence for earlier stellar mass assembly of satellites around Milky Way-mass hosts.

Considering the varying results across simulations, it is possible that differences in the star formation and feedback algorithms for the different simulations could cause differences in the evolution of the SMHM relation with redshift, resulting in changes to (or eliminating) any trend between $M_*/M_{\mathrm{vir, peak}}$ and environment. 
As such, observational tests of the effect of environment on the SMHM may offer a way to compare different sub-grid prescriptions.
Since direct observational measurements of the halo mass are extremely difficult at these masses, in the following section we explore a number of related observational comparisons that would be more feasible in the foreseeable future.

Semi-analytic models and variations on abundance matching have also been used to compare the SMHM for central and satellite galaxies.
The first such student to look for this dependency, \citet{Wang2006}, found no difference between the SMHM for satellite and central galaxies was necessary in their model to reproduce basic statistical properties of galaxies given the Millennium Simulation.
Using a similar technique, however, \citet{Watson2013} found that some distinctions between the central and satellite SMHM relation was necessary to match the galaxy stellar mass function for centrals and the galaxy two-point correlation function for satellites.
Specifically, the stellar masses of satellites were about 10\% higher at z = 0 than central galaxies of the same peak halo mass, which they attribute to slowed stellar growth following infall.
\citet{Behroozi2019} found a similar effect, with sub-L$_*$ satellites having higher $M_*/M_{\mathrm{halo, peak}}$, which they also attribute to slow quenching post-infall.
The data in \citet{Watson2013} extends only down to $10^{11} \Msun$ and in \citet{Behroozi2019} only down to $2 \times 10^{10} \Msun$
These papers also focus on the binary distinction between satellite and central, rather than examining a more continuous environmental effect.
Nevertheless, their results are consistent with ours.

More recently, \citet{Read2017} did find an environmental difference for the SMHM relation when determined through abundance matching, but in the opposite direction observed in our data.
They find that both abundance matching using the `field galaxy' stellar mass function from the Sloan Digital Sky Survey (SDSS) and the direct measurement of isolated dwarf galaxy stellar masses and rotation curves \citep{Read2016} produce similar results for the SMHM relation.
Their data is shown in comparison to ours in Figure~\ref{fig:SMHM}, and is roughly consistent with our isolated dwarf galaxies down to the lower bound of the SDSS stellar mass function ($M_* = 10^7 \Msun$).
They also use abundance matching to compute the SMHM relation for nearby groups using the observed stellar mass function.
This SMHM relation for group galaxies lies to the right of their isolated galaxies, opposite what we see and inconsistent with our data. They attribute the shallower stellar mass function of galaxies in groups to the premature truncation of their star formation by interactions with the host galaxy environment. 
As a result, they argue that $M_*/M_{\mathrm{vir}}$ is lower for satellite galaxies than isolated galaxies, when $M_{\mathrm{vir}}$ is determined through abundance matching.
Further, they argue that the stochastic nature of interactions means that abundance matching is not a reliable technique for determining the halo mass of galaxies in rich environments, 
primarily because 
the assumption of a monotonic relation between $M_*$ and $M_{\mathrm{vir}}$ may break down.
A more accurate comparison of the SMHM for different environments will, therefore, require analysis beyond straightforward abundance matching.

\subsection{Observational Implications}\label{sec:obs}

The relationship between the stellar content of dwarf galaxy halos and the environment they are formed in has consequences both for the properties of observed galaxies and how those observations are interpreted.
In this section, we compare our sample of simulated galaxies to the star formation-related properties of observed dwarf galaxies spanning a range of environments.
We further show how changes in $M_*/M_{\mathrm{vir}}$ with environment can affect the interpretation of those same observations.

\begin{figure}
\begin{center}
\includegraphics[width=0.5\textwidth]{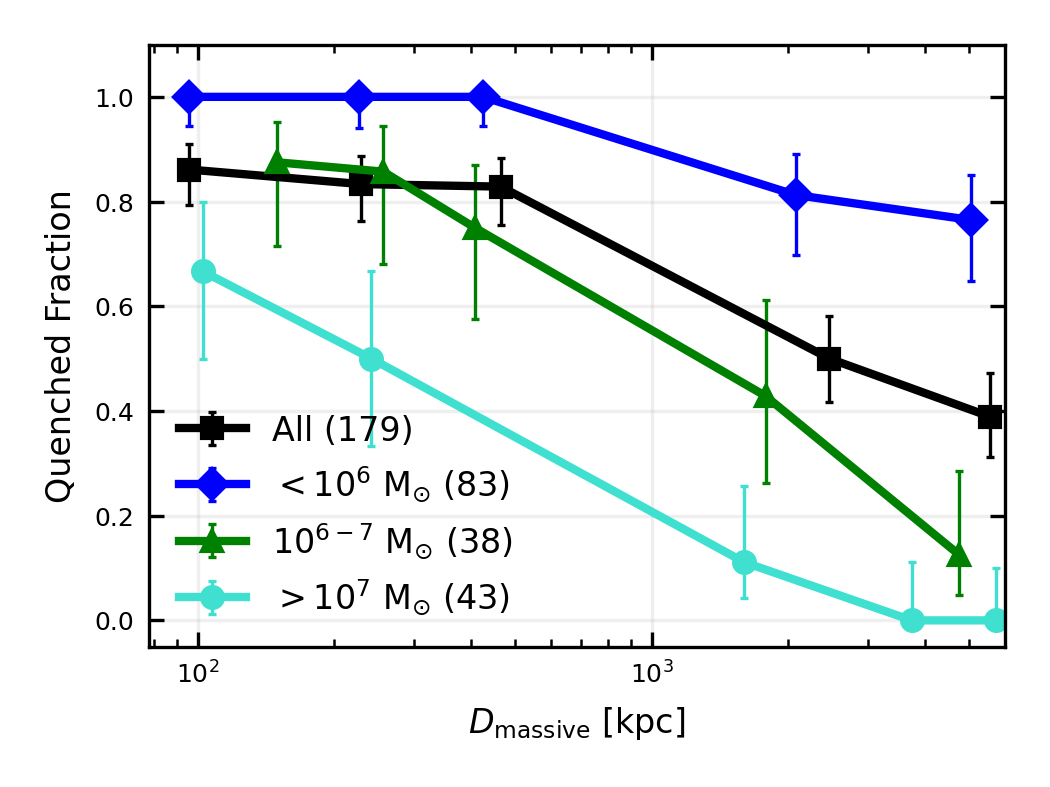}
\end{center}
\caption
{ 
{Fraction of quenched halos versus the distance to the nearest massive} ($M_{vir} > 10^{11.5} \Msun$) galaxy.  Data for the entire population of dwarf galaxies with at least 100 star particles is shown by the black squares. 
The data is further divided by stellar mass, with cyan circles representing ``classical'' dwarfs in the range $10^7$ -- $10^9 \Msun$, green triangles representing dwarfs in the range, $10^6$ -- $10^7 \Msun$, and blue diamonds representing the ultra-faint dwarfs in the range $ < 10^6 \Msun$. 
Within each stellar mass sample, data is binned into quintiles by distance such that each data point represents roughly equal numbers of galaxies.
Error bars represent 68\% uncertainty in the binomial proportion via the Wilson Score Interval \citep{Wilson1927}.
In every stellar mass sample, the fraction of quenched galaxies decreases with distance.
However, the transition from primarily quenched to primarily star forming happens at lower stellar mass for field galaxies.
}
\label{fig:quenchFrac}
\end{figure}

\subsubsection{Quenched Fraction} Observations of the effect of environment on the evolution of dwarf galaxies tend to focus on whether or not they are quenched.
In Figure~\ref{fig:quenchFrac}, we show the fraction of galaxies of different stellar masses that are quenched (defined as having a specific SFR $10^{-11}$ yr$^{-1}$ within the past 200 Myr as measured directly from the simulation) as a function of their $z = 0$ distance to a massive ($M_{vir} > 10^{11.5} \Msun$) galaxy.
As expected, lower-stellar mass galaxies are more likely to be quenched than higher-stellar mass galaxies within the same environment.
Furthermore, across all stellar mass ranges, quenched fraction decreases with distance from a massive galaxy, and this trend persists even for distances greater than 1 Mpc.
These trends are consistent with our previous analysis of the star formation histories and $\tau_{90}$ values as a function of environment.

\citet{Geha2012} similarly identified a strong correlation between distance to a massive (M$_* \gtrsim 2.5 \times 10^{10} \Msun$) galaxy and the observed quenched fraction of dwarf galaxies.
In their analysis, the quenched fraction for dwarf galaxies with $10^8 \Msun < M_* < 10^{9.75} \Msun$ within 1.5 Mpc of a massive galaxy increased for smaller distances, reaching peak quenched fractions between 0.22 and 0.31, depending on the dwarf stellar mass.
Beyond 1.5 Mpc from the host galaxies, no dwarf galaxies with $10^8 \Msun < M_* < 10^{9} \Msun$ were observed to be quenched.
We find the same trends in quenching for our most massive dwarf galaxy sample ($10^7 \Msun < M_* < 10^{9} \Msun$, shown as the circle-marked cyan line in Figure~\ref{fig:quenchFrac}).  
In particular, no galaxy in our sample with $10^8 \Msun < M_* < 10^{9} \Msun$ (42 in total) was quenched beyond 350 kpc from a massive galaxy.
Within 350 kpc ($\sim$ R$_{vir}$), 2/8 of the galaxies with $10^8 \Msun < M_* < 10^{9.75} \Msun$ were quenched, consistent with the findings of \citet{Geha2012}.

For galaxies with $M_* < 10^{7} \Msun$, the environment has a much smaller effect.
Indeed for galaxies with $10^4 \Msun < M_* < 10^6 \Msun$ the quenched fraction is 100\% in all but the largest two distance bins.  
The reduced environmental effect on the quenching of low-mass galaxies may be attributed to the dominance of quenching by the cosmic UV background in this mass range.
Especially in these simulations in which the cosmic UV background is approximated as spatially uniform, the effect on quenching will be independent of environment.
Quenching via reionization is thought to be most likely for galaxies with $M_* \lessapprox 10^6 \Msun$ \citep{Bovill10a}.
This is the same mass range that \citet{Weisz2014} found their reionization candidates, although they also found galaxies in this mass range with extended star formation histories.
Clearly, even in this ultra-faint dwarf regime, the environment still plays some role in our simulations: while no satellites in this stellar mass range are star-forming, seven isolated ultra-faint dwarf galaxies are.
Five of these low-stellar mass star-forming galaxies have late-starting start formation histories and the other two reignited star formation after about 10 Gyrs.
Notably, all of these utlra-faint, star forming galaxies would lie above the WALLABY detection limit of $M_{\mathrm{HI}} \sim 1.9\times10^4\times D[\mathrm{Mpc}]^2 \Msun$, if the distance to the nearest massive galaxy is taken to be, $D$, the distance to the Milky Way \citep{Koribalski2020}.

For galaxies with $M_* > 10^{7} \Msun$, it is notable that we find some enhanced quenching out to $\sim 2 R_{vir}$.
The predominant process by which satellite star formation is quenched is thought to be the removal of gas through ram pressure stripping \citep[e.g.][]{Grebel2003, Mayer2006}. 
The ability of environmental quenching to take place beyond the virial radius is debated in the literature.
While \citet{Geha2012} found enhanced quenching out to 1 Mpc ($\sim 3$ times the virial radius of Milky Way-mass galaxies), \citet{Weisz2015} found no evidence for quenching of Milky Way or M31 satellites beyond 300 kpc at any redshift.  On the other hand, ram pressure out to $\sim 2 R_{vir}$ is routinely found within simulations \citep[e.g.,][]{Bahe2013, Behroozi2014,Fillingham2018}

Quenching might be enhanced outside of $R_{vir}$ if subhalos are ``preprocessed'' within another halo or group prior to their infall to their final host. \citet{Li2008, Slater2013, Bakels2021} all found that a significant fraction of satellites in their simulations were accreted as members of a group.  In host galaxies like the Milky Way, satellites with $M_* > 10^7 \Msun$ typically do not enter quenched \citep{Wetzel2013, simpson_quenching_2018, Akins2021, Samuel2022}, but such interactions prior to accretion may still result in the increased consumption of gas.  However, simulations have indicated weak evidence for enhanced quenching due to preprocessing.
\citet{Wetzel2015} found that a significant ($\sim$25\%) fraction of satellites of Local Group analogs were part of a group prior to accretion.
Nevertheless, in observational analysis informed by these simulations, quenching of lower mass satellites ($10^6 M_*/\Msun < 10^8$ ) happened around or within a couple Gyr after the time of accretion \citet{Wetzel2015}.
\citet{Samuel2022} found that while preprocessing was common (about 40\% of their simulated satellites with $M_* > 10^6 \Msun$ were in low-mass groups prior to accretion) most galaxies with $M_* \gtrsim 10^{6.5} \Msun$ were not quenched prior to infall and of those that did quench prior to infall, the vast majority did so as central galaxies.
Similarly, \citet{Joshi2021} found negligible differences in the pre-accretion star formation histories of preprocessed satellites of $10^{8} \Msun \leq M_{\mathrm{vir}} \leq 10^{12} \Msun$ hosts compared to a mass-matched sample of isolated galaxies.

A second explanation for the presence of quenched galaxies outside of $R_{vir}$ is that they are backsplash galaxies \citep{Gill2005} that have already experienced one pericentric passage within $R_{\mathrm{vir}}$. Simulations show that it is common for galaxies beyond $R_{\mathrm{vir}}$ to have been previously within it.
For example, in an analysis of dark matter-only simulations, \citet{Bakels2021} found that roughly half of previously-accreted subhalos reside beyond $1.2 R_{\mathrm{vir}}$ at $z = 0$. Indeed, \citet{simpson_quenching_2018} estimates that some 40\% percent of all dwarf galaxies within 1 Mpc of a Milky-Way mass host but outside of its virial radius at $z = 0$ are backsplash galaxies.
\citet{Teyssier2012} estimated 13\% within 1.5 Mpc of the Milky Way are backsplash galaxies and \citet{Buck2019} estimated 50-80\% within 2.5 R$_{\mathrm{vir}}$. 
\citet{Wetzel2014} used an analysis of SDSS galaxies combined with orbital information from dark-matter only simulations to argue that the quenching of backsplash galaxies during their time within a MW-mass host halo can account for enhanced quenched fractions out to $5 R_{\mathrm{vir}}$.
\citet{simpson_quenching_2018} found that backsplash systems  had quenching patterns similar to those satellites within $R_{\mathrm{vir}}$.
Therefore, any study of quenched fractions as a function of distance must consider the presence of backsplash systems.

\begin{figure}
\begin{center}
\includegraphics[width=\linewidth]{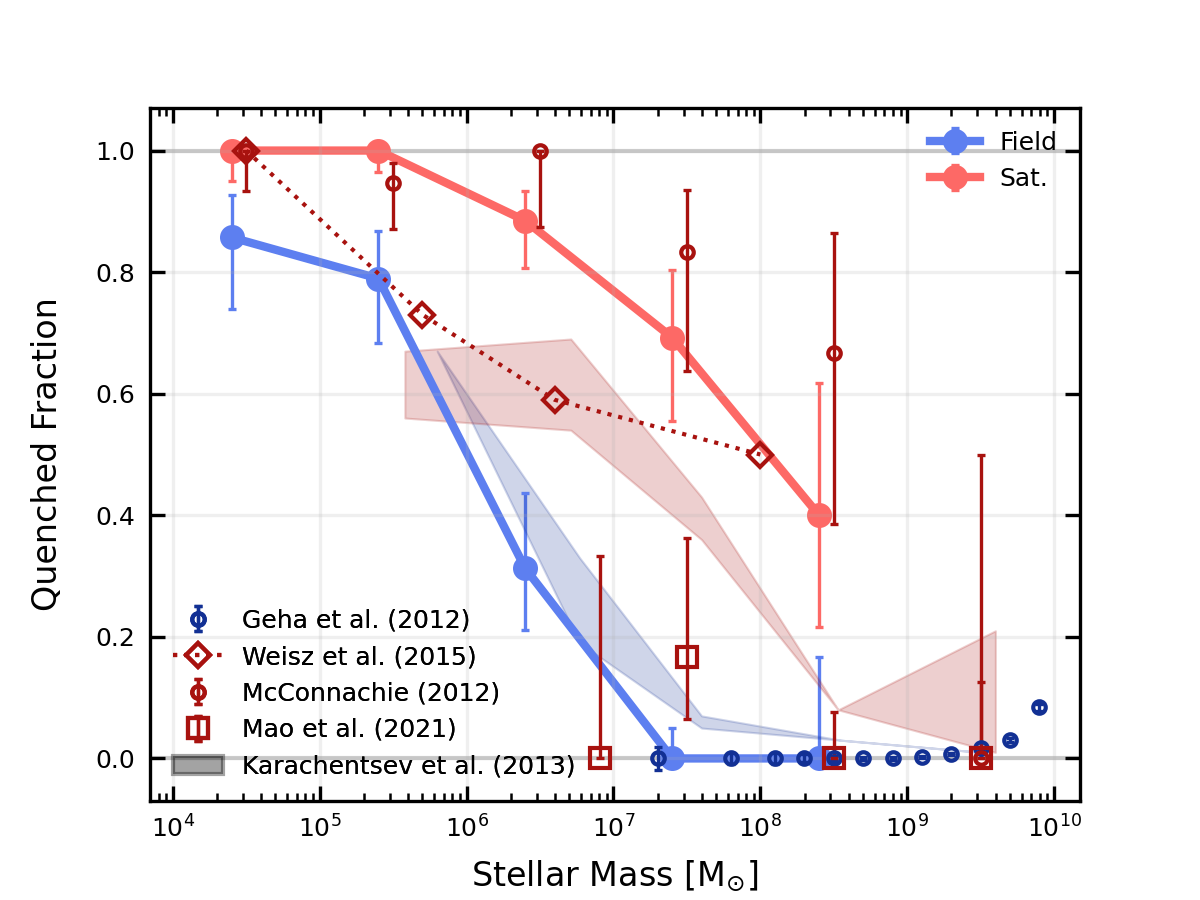}
\end{center}
\caption
{ 
A comparison between the quenched fraction as a function of stellar mass for simulated field galaxies (blue-filled circles) and simulated satellite galaxies (red-filled circles).
Error bars represent 68\% uncertainty in the binomial proportion via the Wilson Score Interval \citep{Wilson1927}.
The blue-filled regions \citep{Karachentsev2013} and open blue circles \citep{Geha2012} are both observational comparisons for field galaxies.
The red-filled regions \citep{Karachentsev2013}, red open diamonds \citep{Weisz2015}, red open circles \citep{McConnachie2012}, and red open squares \citep{Mao2021} are observational comparisons to satellite galaxies. 
The transition from primarily quenched to primarily star forming happens at lower stellar mass for field galaxies, and even in the lowest stellar mass bin, the simulated field galaxies are less likely to be quenched than satellites.
}
\label{fig:obscomp}
\end{figure}

Figure~\ref{fig:obscomp} shows observational comparisons to the quenched fraction of satellites of MW-mass hosts as a function of stellar mass.
In general, observations of satellite galaxies \citep[e.g.][]{Weisz2015,McConnachie2012,Karachentsev2013}  have found increasing quenched fractions with lower stellar mass, and uniformly higher quenched fractions than field galaxies of similar stellar masses.
This trend, however, is complicated by data from the recent Satellites Around Galactic Analogs survey \citep{Mao2021}, which found very low quenched fractions for galaxies with stellar masses $> 10^{7} \Msun$.
Nevertheless, data from our simulations is generally consistent with observations of the Local Group.
We refer the reader to \citet{Akins2021} for a more extensive analysis of the comparison between the near-mint {\sc D.C.\,Justice League} satellites and observations.

\begin{figure}
    \centering
     \includegraphics[width=\linewidth]{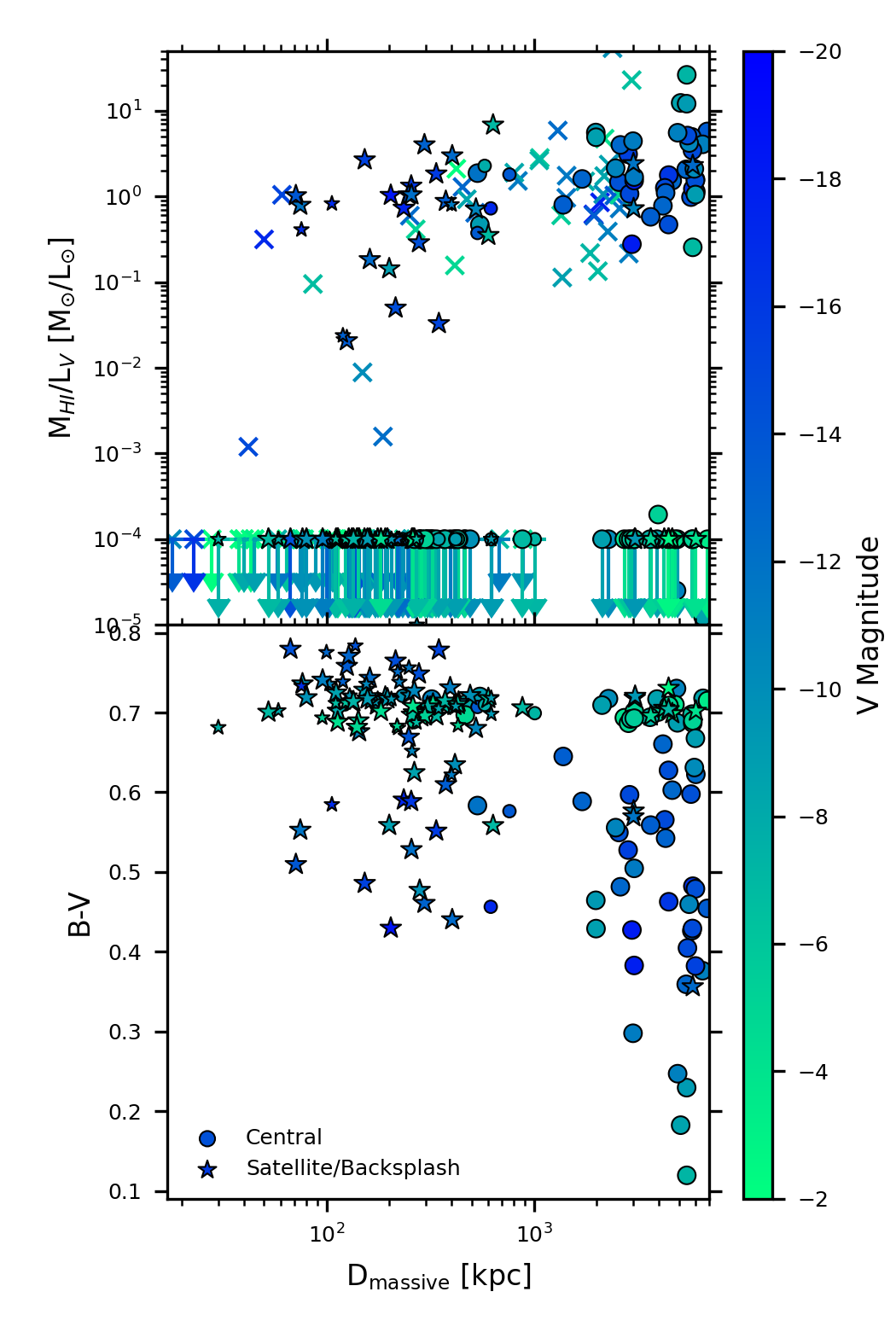}
    \caption{Top: Ratio of \textsc{H\,i} to $V$-band luminosity vs. distance to nearest massive galaxy, with points colored by $V$-band magnitude. Downward arrows show upper-limits to the gas fraction. Bottom: Ratio of $B-V$ color vs. distance to nearest massive galaxy, with points colored by $V$-band magnitude. In both panels, simulated galaxies are shown as stars and circles. In the top panel, $\times$ symbols represent observed Local Group from \citet{McConnachie2012}, with the distance being the minimum distance to either the Milky Way or M31. Our \textsc{H\,i} data matches well with observations, continuing the trend past 3000 kpc.}
    \label{fig:color_fHI_dist}
\end{figure}

\subsubsection{Gas Content and Color}

\begin{figure*}
    \centering
    \includegraphics{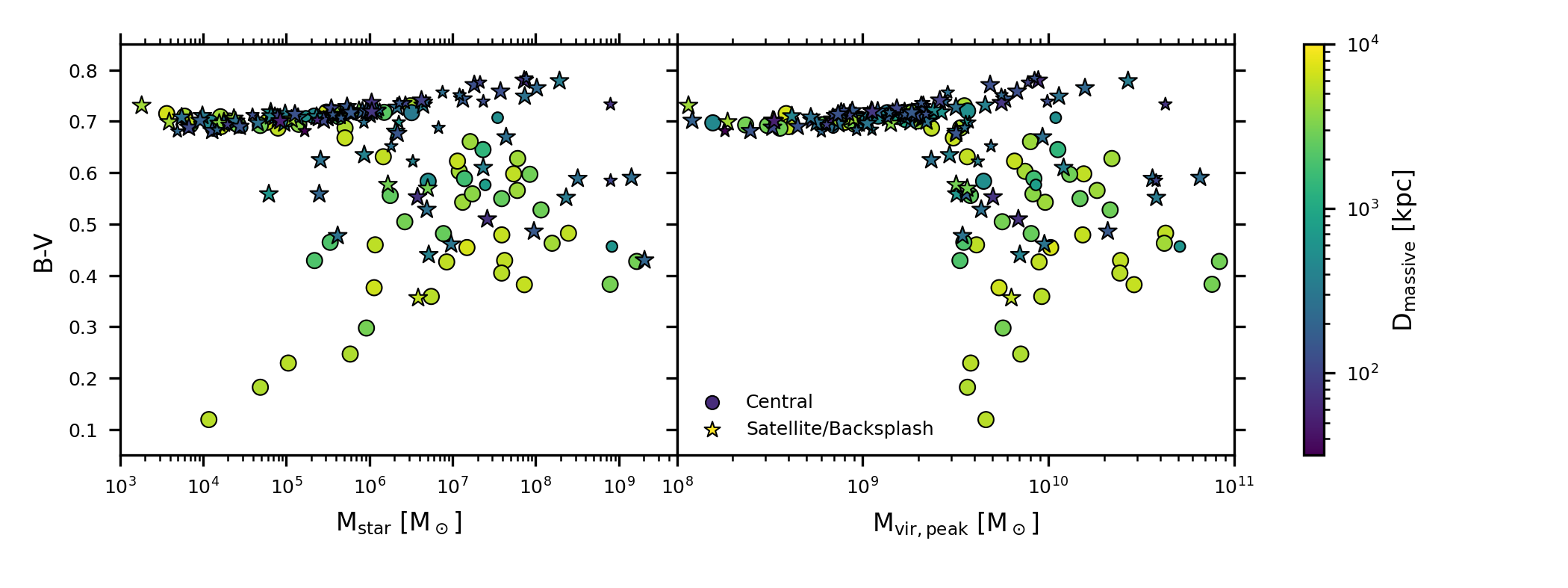}
    \caption{$B-V$ color of simulated dwarf galaxies versus their $z = 0$ stellar mass (left panel) and peak virial mass (right panel). The color of the points represents the current distance to a massive galaxy.
    While isolated dwarf galaxies may be blue, galaxies with $M_{\mathrm{vir, peak}} \lesssim 2 \times 10^9 \Msun$ are uniformly red.
    Therefore, the isolated, extremely low-stellar mass blue galaxies apparent in the left panel inhabit moderate-mass dark matter halos.}
    \label{fig:BV_mass}
\end{figure*}
While the quenched fraction as a function of distance offers a binary examination of the galaxies' star formation, the color as a function of distance can offer a more nuanced portrait.
Along with color, \textsc{H\,i} is an excellent indicator of star formation; halos without \textsc{H\,i} will not be able to form stars without more gas inflow. 
The upper panel of Figure \ref{fig:color_fHI_dist} shows the ratio of \textsc{H\,i} mass to $V$-band luminosity compared to the distance to the nearest host. 
This graph is a recreation of the same measurement from \citet{McConnachie2012}, whose information is included as $\times$ symbols. 
Arrows at the bottom represent halos that have close to or no \textsc{H\,i} mass ($<10^{-4}$ $M_{\odot}/L_{\odot}$). 
Consistent with McConnachie's data and the more recent compilations from \citet{Spekkens2014} and  \citet{Putman2021}, there is an increase in \textsc{H\,i} mass with greater distance from massive halos. 
Additionally, within a virial radius of a Milky Way-mass halo observations from \citet{Karunakaran2022} have found a transition magnitude range from $-10 \geq M_V \geq -14$ wherein galaxies may be either gas-rich or gas-poor, consistent with our results.

The bottom panel of Figure \ref{fig:color_fHI_dist} displays the $B-V$ color of the galaxies as a function of distance from a massive galaxy.
While a large number of red ($B-V \geq 0.7$) galaxies exist at all distances, the number of galaxies with $B-V < 0.6$ increases with distance from a massive galaxy. 
Furthermore, the average color of galaxies with $B-V < 0.6$ is bluer for increased distances.
Some of this color trend is driven by the five late-star forming dwarf galaxies (see figure~\ref{fig:sfh}), which are visible on the graph as the bluest galaxies.
However, even if the delay in star formation for these galaxies is deemed to be numerical rather than physical, the trend between color and environment remains.
This panel is the observational equivalent to the lower left panel of Figure~\ref{fig:taus}.
In both instances, greater isolation corresponds to a weighting of star formation to later times and bluer colors.

Figure \ref{fig:color_fHI_dist}  shows a significant number of simulated halos beyond 1000 kpc with no \textsc{H\,i}, presumably quenched during reionization.
Observational proxies for the most distant halos with $V$ magnitudes dimmer than -7 are generally unavailable \citep{Simon2019}.
One notable exception is Tucana B, a quenched ultra-faint dwarf galaxy located 1.4 Mpc away and likely quenched during reionization \citep{Sand2022}.
Similarly, the handful of isolated galaxies with $B-V < 0.4$ in our simulations are as of yet mostly undetectable.
These galaxies all have late-starting star formation histories and without observations, we cannot say whether their star formation histories are physical numerical.
Recently, though, \citet{Janesh2019} used WIYN follow-up imaging of ALFALFA sources to detect five gas-rich ultra-faint dwarf galaxy candidates with \textsc{H\,i} gas masses between $2 \times 10^4$ and $3 \times 10^6 \Msun$, stellar masses from $4 \times 10^2$ to $4 \times 10^5 \Msun$, and distances between $\sim 350$ kpc to $\sim 1.6$ Mpc.
While these candidates differ from ours in not showing evidence for young stellar populations, the similarity of distance, \textsc{H\,i}-mass, and stellar mass is intriguing. 
Similar observational techniques may be able to detect  gas-rich, low-stellar mass galaxies star-forming galaxies similar to those seen in these simulations.
In particular, Apertif \citep{vanCappellen2022} and WALLABY \citep{Koribalski2020} will enable a census of much lower \textsc{H\,i}-mass galaxies within the Local Volume than had previously been available, and all seven of our isolated low-luminosity, star forming galaxies (including those with both late-starting and reigniting star formation histories) would lie above their detection limit.

Observational analysis of dwarf galaxy star formation necessarily focuses on trends as a function of stellar mass, rather than $M_{\mathrm{vir, peak}}$ or even $M_{\mathrm{vir}}$.
The left-hand panel of Figure~\ref{fig:BV_mass} summarizes these trends by showing color as a function of stellar mass for galaxies at different distances. 
As seen in the previous analysis in this section, isolated galaxies are on average bluer and more likely to be star forming than satellite galaxies of similar masses.
This trend holds whether or not the five late-starting star forming galaxies with $B-V < 0.3$ and occupying moderate-mass dark matter halos are included.
As previously shown in the analysis of the SMHM relation, the relationship between $M_*$ and $M_{\mathrm{vir, peak}}$ also depends on environment.
Therefore, as shown in the right-hand Figure~\ref{fig:BV_mass}, trends between recent star formation and environment dramatically change when $M_{\mathrm{vir, peak}}$ is considered rather than $M_*$.
Galaxy color versus $M_*$ shows a bimodal relationship across all stellar masses with the bluest galaxies being more likely to be isolated.
When galaxy color is shown versus $M_{\mathrm{vir, peak}}$, though, the separation between satellite and field galaxies only exists for galaxies with $M_{\mathrm{vir, peak}} \gtrsim 2 \times 10^9 \Msun$.
Below this mass, all galaxies have similar red colors, i.e., they are quenched.
This similar behavior for field and satellite dwarf galaxies is the result of quenching by the spatially uniform cosmic UV background in our simulations during reionization.

\subsection{Resolved Star Formation Histories}

\begin{figure}
    \centering
    \includegraphics[width=0.99\columnwidth]{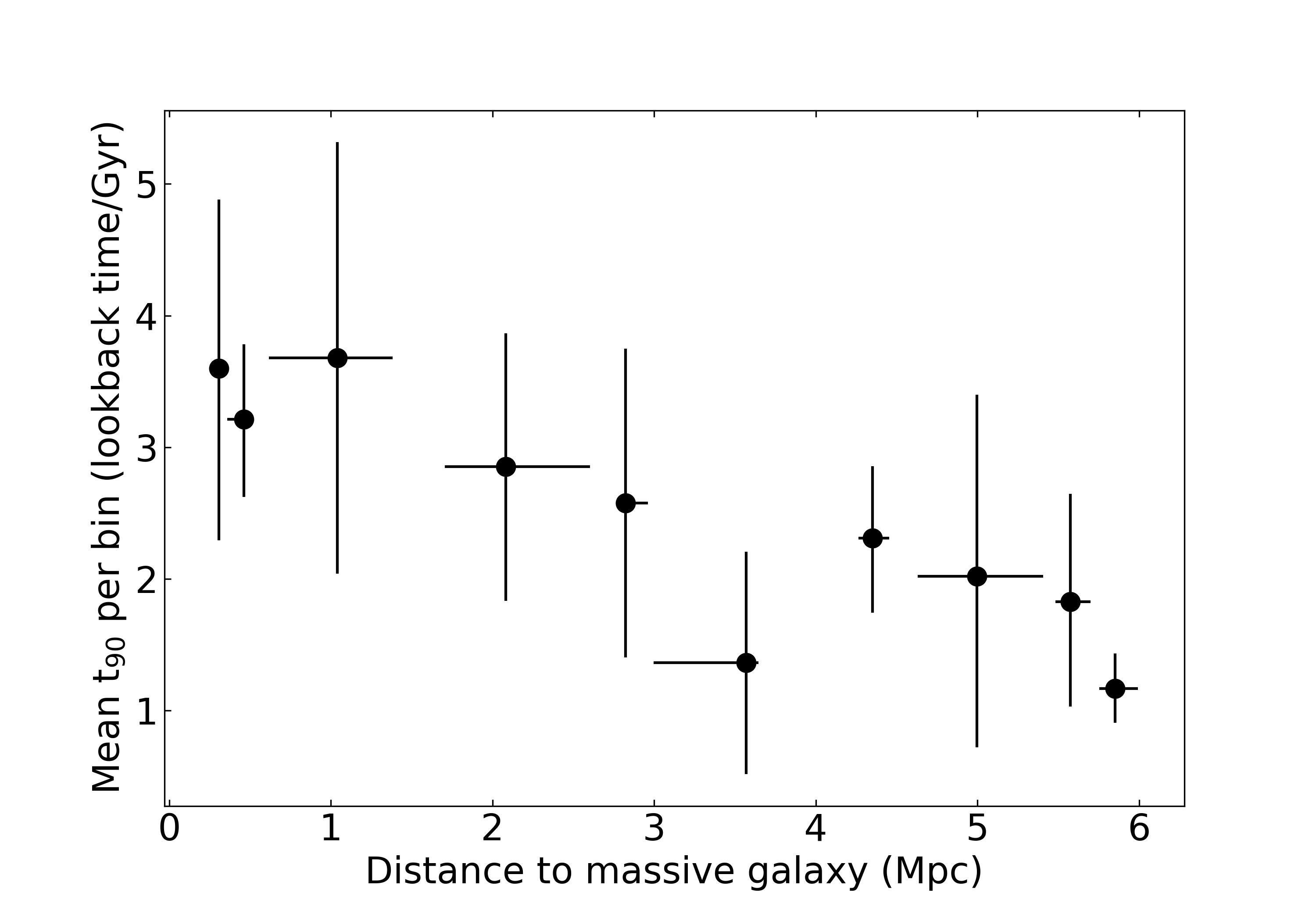}
    \caption{Mean $\tau_{90}$ for field dwarfs as a function of distance to a massive galaxy.  Simulated galaxies with $M_* > 10^6 M_{\odot}$ and $\tau_{90}$ lookback times within the last 6 Gyr are included.  Bin size is set so that there are an equal number of galaxies (four) in each bin.  Error bars on $\tau_{90}$ reflect the 1$\sigma$ standard deviation of the values in that bin.  Data points along the x-axis represent the center of each bin, with the error bars showing the minimum and maximum distances of the galaxies found in each bin.  There is a clear trend for galaxies further away from a massive galaxy to have a more recent $\tau_{90}$. }
    \label{fig:t90dist}
\end{figure}

In Figure~\ref{fig:smassdist} we showed that $\tau_{90}$, the time at which star formation is 90\% complete, varies strongly with distance to a massive galaxy.  Integrated-light-based star formation tracers such as UV or H$\alpha$ only probe the most recent star formation ($\lesssim 100$ Myr), but observations of resolved stellar populations allow the derivation of $\tau_{90}$ by fitting the stellar populations seen in color--magnitude diagrams.  In this section, we make predictions for $\tau_{90}$ as a function of environment that could potentially be examined by current data. 

The earliest star formation in galaxies can only be traced using resolved stellar populations that reach below the oldest main-sequence turnoff \citep[e.g.,][]{Cole2007, Cole2014, Albers2019}.  Most resolved star data does not reach sufficient depth, however, to trace the oldest main-sequence turnoff \citep[e.g.,][]{Weisz2011, Weisz2014}.  In Figure \ref{fig:t90dist}, we restrict our simulated galaxy sample to those with $\tau_{90}$ within the last 6 Gyr, because the error bars on age are generally too large at older times for most existing resolved star data to accurately determine.  We also limit our sample to field galaxies with $M_* > 10^6 M_{\odot}$.  These cuts remove galaxies that have been directly impacted by reionization and reduce the contribution from galaxies that have experienced ram pressure stripping.  The result is a galaxy sample better matched to current observations of field galaxies with resolved stellar populations. 

Figure \ref{fig:t90dist} bins this observationally aligned sample of simulated galaxies as a function of distance to a massive neighbor, again defined to have $M_{vir} > 10^{11.5}$ M$_{\odot}$.  The bin sizes have been determined by requiring an equal number of galaxies in each bin (four galaxies).  The $x$-axis data points represent the middle of each bin, while the $x$-axis error bars reflect the smallest and largest distances within each bin.  The $y$-axis is the mean $\tau_{90}$ of the galaxies in each bin, though this time they are plotted as lookback time, unlike in earlier plots.  The $y$-axis error bars reflect the standard deviation within each bin.  

Figure \ref{fig:t90dist} highlights a potentially observable trend: there is a clear prediction for galaxies to have a more recent $\tau_{90}$ as distance to a massive neighbor increases.  This trend may potentially be observable with current archival HST data, but also motivates future observations with JWST that can probe to similar depths as HST but to a greater distance from the Milky Way.  Such observations have the potential to directly test the accuracy of our galaxy formation model.

\section{Conclusion}
In this work, we compare the evolution of simulated low-mass galaxies formed within different environments to $z = 0$.
Specifically, we compare suites of low mass galaxies formed in close proximity to Milky Way-mass galaxies (the {\sc D.C.\,Justice League} simulations) to those formed in more isolated environments, comparable to the Local Volume (the {\sc Marvel-ous Dwarfs}).
By comparing galaxies of similar peak virial mass and using the distance of closest approach to a massive ($M_{\mathrm{vir}} > 10^{11.5} M_\odot$) galaxy as a measurement of environment, we found the following differences.
\begin{enumerate}
    \item Low-mass galaxies formed in more isolated environments tend to have lower stellar masses than galaxies of similar peak halo masses formed in denser environments. Applying an extra sum squares F-test provides a $p$ value of $7.5\times10^{-9}$ for the significance of this result. 
    
    \item Low-mass galaxies formed in more isolated environments tend to have higher gas fractions (defined as either $M_{\mathrm{HI}}/(M_{\mathrm{HI}} + M_{*})$ or $M_{\mathrm{HI}}/L_V$) and are less likely to be quenched. However, galaxies with $M_{\mathrm{vir, peak}} \lesssim 2 \times 10^{9} M_\odot$, are uniformly quenched, regardless of environment, indicating the dominant role of the spatially uniform UV background in our simulations for quenching within this mass range.
    
    \item Low-mass galaxies formed in more isolated environments show evidence of slower mass assembly. The time within which 50\% and 90\% of the maximum virial mass is assembled ($\tau_{\mathrm{50, vir}}$ and $\tau_{\mathrm{90, vir}}$, respectively) occurs later with increasing distance from a massive galaxy. Additionally, the time at which the virial mass peaks also occurs later (and is more likely to be the current age of the Universe) with increasing distance from a massive galaxy. This change in the time of peak virial mass cannot only be due to the tidal stripping of satellite galaxies by a massive companion, as central galaxies also show this effect when considered alone.
    
    \item Low-mass galaxies formed in more isolated environments have star formation histories weighted toward later times. Similar to $\tau_{\mathrm{50, vir}}$ and $\tau_{\mathrm{90, vir}}$, $\tau_{90}$ for galaxies with $M_{\mathrm{vir, peak}} > 2 \times 10^{9} M_\odot$ (i.e., unlikely to have been quenched by reionization) rises with distance. As a result, more isolated galaxies also have bluer colors. 

    \item In contrast to the differences in the stellar and \textsc{H\,i} masses when considered individually, the total disk baryonic mass ($M_{\mathrm{HI}} + M_{*}$) shows a much reduced environmental dependency. In effect, more isolated galaxies have similar $z = 0$ disk baryonic content but have been less efficient at converting those baryons into stars. This premise that more isolated galaxies are slower to convert baryons into stars is consistent with the later star formation histories of isolated galaxies, although the analysis is complicated by the ability of ram pressure stripping to remove potentially star forming gas from satellite galaxies \citep[e.g.,][]{murakami_interaction_1999, mayer_simultaneous_2006, slater_mass_2014, bahe_star_2015, simpson_quenching_2018}.

\end{enumerate}

Notably, {\em none of the trends enumerated above was limited to contrasting satellite and backsplash galaxies to central galaxies}.
Indeed, when central galaxies alone are considered, the above trends remain, albeit across a smaller dynamical range.
The fact that these differences persist even for central galaxies indicates that they cannot be solely attributed to direct interactions with a more massive host galaxy.
Instead, the environment must also impact the evolution of these galaxies either through differences in the timescale of gravitational collapse or in the frequency of dwarf-dwarf interactions--two effects that in hierarchical galaxy formation are not completely distinct.
Additionally, increases in gas-phase metallicity in denser environments could increase rates of gas cooling and star formation but is unlikely to have affected the overall assembly history of the galaxies.
(As previously noted, the UV-background is spatially uniform in these simulations, and so cannot be the source of differences in these simulations, although spatial differences may have effects in the actual Universe.)

Overall, our results still require additional work to understand the underlying physical mechanisms that drive these trends.  Despite that, we find a clear signature of the dependency of star formation on environment within the Local Volume, leading to strong trends in star formation histories and gas content that can be directly tested.  With the likely discovery of hundreds of dwarf galaxies within the Local Volume in the next few years (via their stellar content in LSST, or via their \textsc{H\,i} content in WALLABY), we will have the exciting opportunity to confront dwarf galaxy formation models.

\section{Acknowledgements}\label{sec:acknowledge}
We appreciate the feedback from the anonymous referee.
We are grateful to Shonda Kuiper for her assistance in the statistical analysis.
This work was supported by the U.S. NSF under CAREER grant AST-1848107. 
H.~A.~ and L.~C.~were supported by the Grinnell College Mentored Advanced Project program.
A.~M.~B.~was partially supported by NSF grant AST-1813871. 
Resources supporting this work were provided by the NASA High-End Computing (HEC) Program through the NASA Advanced Supercomputing (NAS) Division at Ames Research Center.


\bibliography{satelliteProperties}{}
\bibliographystyle{aasjournal}

\end{document}